\documentclass[12pt]{article}
\usepackage{amsmath}
\numberwithin{equation}{section}

\usepackage{color}
\usepackage{graphicx}
\usepackage{float}
\usepackage{amsfonts}
\usepackage{bm}
\usepackage{amsmath}
\usepackage{amscd}
\usepackage{amssymb,lscape}
\usepackage{epsfig}

\newcommand{\bmat}{\left(\begin{array}}
\newcommand{\emat}{\end{array}\right)}

\def\gtrsim{\mathrel{\raise.3ex\hbox{$>$\kern-.75em\lower1ex\hbox{$\sim$}}
}
}

\def\ap{\alpha^{\prime}}

\def\-{\hphantom{-}}

\def\s2{\frac{1}{\sqrt2}}

\def\beq{\begin{equation}}
\def\eeq{\end{equation}}
\def\beqa{\begin{eqnarray}}
\def\eeqa{\end{eqnarray}}

\def\mg{m_{3/2}}
\def\mg2{m^2_{3/2}}

\def\Dsl{\,\raise.15ex\hbox{/}\mkern-13.5mu D} 

\def\be{\begin{equation}}
\def\ee{\end{equation}}
\def\bea{\begin{eqnarray}}
\def\eea{\end{eqnarray}}

\newcommand{\nn}{\nonumber}

\makeatletter
\@addtoreset{equation}{section}
\makeatother

\begin{document}
\pagestyle{plain}
\begin{titlepage}
\begin{center}
  \LARGE{
    Probing the String Winding Sector  
\\[1mm]}
\large{\bf  Gerardo Aldazabal${}^{a,b}$, Mart\' in
Mayo${}^{a,b}$,  Carmen Nu\~nez${}^{c,d}$
 \\[4mm]}
\small{
${}^a${\em Centro At\'omico Bariloche,} ${}^b${\em Instituto Balseiro
(CNEA-UNC) and CONICET.} \\[-0.3em]
{\em 8400 S.C. de Bariloche, Argentina.} \\
[0.4cm]}
${}^c${\em  Instituto de Astronom\'ia y F\'isica del Espacio
(CONICET-UBA) and\\
${}^e$Departamento de F\' isica, FCEN, Universidad de Buenos Aires\\
C.C. 67 - Suc. 28, 1428 Buenos Aires, Argentina}\\

\end{center}
E-mail:\,{aldazaba@cab.cnea.gov.ar},\,{martin.mayo@ib.edu.ar},\,{carmen@iafe.uba.ar}

\vspace{0.5cm}

 {\bf Abstract}:
We probe a slice of the massive winding sector of bosonic string theory from 
toroidal compactifications of Double Field Theory (DFT). This string subsector 
corresponds to states containing one left and one right moving 
oscillators. 
We perform a generalized Kaluza Klein compactification of DFT on generic  
$2n$-dimensional toroidal constant backgrounds and show that,  up to third order 
in fluctuations, the theory coincides with the corresponding  effective  theory 
of the bosonic string 
 compactified on $n$-dimensional toroidal constant backgrounds, obtained from 
three-point amplitudes.
The comparison between both theories is facilitated 
by noticing that generalized diffeomorphisms in DFT allow to fix generalized 
harmonic gauge conditions that help in identifying the physical degrees of 
freedom.
These conditions manifest as  conformal anomaly cancellation requirements on the 
string theory side.
The explicit expression for the gauge invariant effective action
containing the physical massless sector (gravity+antisymmetric+gauge+ scalar 
fields)
coupled to towers of generalized Kaluza Klein  massive states 
(corresponding to compact momentum and winding modes) is found. The action 
acquires 
a very compact form when written in terms of fields carrying $O(n,n)$  indices, 
and is 
explicitly T-duality invariant.
The global algebra associated to the generalized Kaluza Klein compactification is 
discussed.

\vspace{1cm}

\today


\end{titlepage}


\begin{small}
\tableofcontents
\end{small}

\newpage\section{Introduction}

Many amazing properties and symmetries of string theory can be tracked down to 
the extended nature of the strings. 
In particular, the presence of an antisymmetric tensor $B_{\hat\mu\hat\nu}$ in the 
spectrum is expected because,
 being one dimensional,  the string directly couples to it. 
 Actually, a distinctive feature of all string 
theories is that, besides the metric $g_{\hat\mu\hat\nu}$, the gravitational sector 
also 
includes the Kalb-Ramond field  $B_{\hat\mu\hat\nu}$ and a scalar dilaton $\phi$, with 
extended 

\begin{equation}
S=\frac{1}{2 \kappa^2}\int d^Dx\, \sqrt{-g}\, e^{-2\phi}\left(R\, +\, 
4\partial_{\hat\mu}\phi\partial^{\hat\mu}\phi\, -\, 
\frac{1}{12}H_{\hat\mu\hat\nu\hat\lambda}H^{\hat\mu\hat\nu\hat\lambda}\right)\, ,\label{eh}
\end{equation}
where $H_{\hat\mu\hat\nu\hat\lambda}\equiv \,\partial_{[\hat\mu}B_{\hat\nu\hat\lambda]}$. 
The occurrence of this universal gravitational sector   is ultimately due to 
the fact that NS-NS massless fields are constructed from the tensor product of one
left and one right moving oscillators,  transforming in the fundamental 
representation of the 
$D$-dimensional Lorentz 
group $ SO(1,D-1)$, and hence accounting for the degrees of freedom of 
$g_{\hat\mu\hat\nu}$, $B_{\hat\mu\hat\nu}$ and $\phi$ according to the decomposition
\begin{equation}
{D}^2=\left(\frac{{D}({D+1})}{{2}}-{1}
\right)\, \oplus\,  \frac{{D}({D-1})}{{2}}\, \oplus\, 
{1}\, .
\end{equation}

   If the 
space is compact, the closed string can  wind around non-contractible cycles, 
leading to the so-called winding states. 
Again, from the world sheet point of view, these states are created by vertex 
operators involving both 
coordinates associated with momentum excitations and dual coordinates 
associated 
with winding excitations or, equivalently,  left and right moving coordinates.

 The  
presence of winding and momentum modes underlies 
T-duality, a genuine stringy feature, which manifests itself by connecting the 
physics of strings defined on geometrically very different 
backgrounds and  give rise to enhanced gauge symmetries at specific points of 
the compact space. Indeed, T-duality implies that $n$-dimensional toroidal 
backgrounds of closed string theory related by
 the non-compact group $O(n,n,{\mathbb  Z})$ are physically equivalent.
 This duality appears as a continuous global $O(n,n, {\mathbb R})$
symmetry   in the Kaluza-Klein (KK) toroidal compactification
 of the corresponding low energy effective gravity theory (\ref{eh}),  if only 
the massless modes are kept.
 Once the massive KK modes are taken into account, the continuous symmetry is 
broken. 

Double Field Theory (DFT)  aims at incorporating these stringy features, and
in particular  information about winding,  into a field theory
 \cite{Siegel:1993xq}-\cite{review}. 
Inspired by  string compactification on tori, DFT is formulated on a doubled 
configuration space, with 
coordinates $\mathbb{X}^{\cal M}=(\tilde  x_{\hat\mu},\, x^{\hat\mu})$, where  
new  
coordinates $\tilde x_{\hat\mu}$,
  conjugate  to  windings,  are added to the standard coordinates  
$x^{\hat\mu}$, 
conjugate to momenta. Here
 ${\cal M}=0,\dots, 2D-1$ and $\hat\mu=0, \cdots, D-1$.
  A manifestly $O(D,D)$ invariant action is then constructed on the doubled 
space, in which the 
global $O(D,D)$ symmetry is linearly realized.
An interesting feature of DFT is that the metric $g_{\hat\mu\hat\nu}$ and 
antisymmetric tensor 
$B_{\hat\mu\hat\nu}$ fields
can be incorporated into a unique field, the so-called generalized metric,  
transforming as a tensor of the $O(D,D)$ group.

 DFT has local invariances that are well defined only if 
consistency constraints are satisfied. 
A solution to these 
constraints is the so called section condition, which effectively leads to the 
elimination of half of the coordinates. Under this solution and in the frame in 
which the fields do not depend on 
$\tilde x_{\hat\mu}$, the DFT action reduces to  (\ref{eh}) and the generalized 
infinitesimal transformations reduce to the standard diffeomorphisms and
gauge transformations of $B_{\hat\mu\hat\nu}$ that leave (\ref{eh}) invariant.

Even if the original
motivation is lost when choosing the section condition, DFT  still provides 
an interesting tool for understanding underlying symmetries of string theory. 
In 
particular, it  shares the basic features 
of Generalized Complex Geometry \cite{Hitchin:2000jd, 
gualtieri} 
(both frameworks are based on an ordinary, undoubled, manifold) and the 
$2D$-dimensional tangent bundle of the doubled space is an extension of the 
$D$-dimensional tangent-bundle of ordinary spacetime by its cotangent bundle, 
with $B_{\hat\mu\hat\nu}$ 
parametrizing the structure of the fibration. Actually,  some 
distinctive ingredients of string theory, like $\ap $ corrections, have been 
recently 
incorporated in these formulations \cite{nos, el}.

Other  solutions to the constraint equations are provided by 
generalized Scherk-Schwarz compactifications \cite{Aldazabal:2011nj, 
Geissbuhler:2013uka}. It is worth noticing that 
Scherk-Schwarz compactifications of DFT give rise to  all the  gaugings of 
gauged supergravity theories (not obtainable from compactifications of 
low energy effective supergravities) allowing for a geometric 
interpretation 
of all of them \cite{dfmr}, albeit in a double space. In this framework, the 
doubled 
coordinates enter in a very 
particular way through the twist matrix, which gives rise to the constant 
gaugings.

While  winding modes are essential for T-duality, they are not truly present 
and  their role is not 
evident  in these approaches. Clearly, to probe the winding sector requires to 
relax the section 
condition. Moreover,
in toroidal string compactifications, winding states are 
massive for generic tori. Therefore, understanding the role of winding modes 
implies 
facing  the massive sector of the theory and  consequently dealing with an 
infinite number of physical states,
with different spins and mass scales.
However, at specific points of the compact space, some winding states become 
massless 
and an effective theory containing only massless states and enhanced gauge
symmetry emerges. This scenario appears particularly suitable to identify 
the explicit part played by windings and a DFT description of the massless 
winding sector 
of bosonic string theory compactified on a circle was suggested in 
\cite{aimnr}. 

In the present  work we  propose a way  to probe a slice of  the massive 
 winding sector of bosonic string theory in an organized fashion.
Namely, we consider compactifications of DFT  on generic double
 tori\footnote{See \cite{hs} for previous work on this subject.}. The
generalized dilaton and metric fields of DFT contain
bosonic string states constructed with one left and one right moving 
oscillators, and therefore we concentrate on this
sector of the string spectrum.
Even if the bosonic string is ill defined, due to the presence of tachyons, 
we will use it as a reference since string computations 
are simpler to deal with. However, for the sector we are interested in here, 
similar 
reasoning would apply for the heterotic or Type IIB string theories.  

The comparison between DFT and string theory is done by  expanding the 
generalized
fields around a generic toroidal background 
with constant dilaton and two-form field\footnote{Expansions around generic 
backgrounds have been performed in \cite{hm}}. We then expand the DFT action up to  
third order 
in fluctuations  around the constant background and contrast the result with the 
corresponding string theory 
three point amplitudes.

As a first outcome of the calculations, we find that both the DFT and string 
spectra containing Kaluza-Klein (KK) momenta and windings
coincide as long as  a ``level matching''  constraint (LMC) is imposed on the 
mode expansion of the DFT fields.
Furthermore, we show that the compactified DFT action (up to this order in fluctuations) 
is invariant under generalized gauge transformations generated by a generalized Lie 
derivative,  if the LMC is imposed.
This gauge invariance allows to choose  a generalized harmonic 
 gauge which provides a  convenient ``gauge fixing'', as it imposes 
conditions on massless and massive states that can be easily identified with 
conformal anomaly cancellation conditions on the vertex operators creating 
these states in string theory. 
Using these conditions, we then show that  cubic vertices in the  DFT
action can be  reproduced by three point amplitudes in string theory. 
Actually, DFT appears to provide a straightforward way of organizing these 
amplitudes in an effective T-duality invariant field theory.
We obtain an explicit expression for the gauge invariant effective action
containing the physical massless sector (gravity+antisymmetric+gauge+scalar 
fields)
coupled to towers of generalized Kaluza Klein (GKK) massive states 
(corresponding to  compact momentum and winding modes).

The article is organized as follows. In Section 2 we present some basic 
introduction to DFT. We write the DFT action in a generalized Einstein frame 
and 
 fix the gauge freedom in terms of
generalized harmonic coordinates. 
In Section \ref{sec:Perturbative DFT} we perform the expansion of the 
generalized fields  in  
fluctuations around a 
constant generic background, we discuss the gauge fixing conditions and  
carry out
a GKK decomposition of the fields. In Section 
\ref{sec:Toroidal compactification} we 
consider 
the mode expansion of the fields on a double torus with constant background 
fields.
 We identify massless and massive states and  examine the generalized harmonic 
 gauge equations to distinguish physical states and Goldstone like states. 
 The analysis of the cubic interaction terms in the 
effective action and the identification of unbroken  symmetries is also 
performed. Finally the resulting gauge invariant action in $d$ lower dimensions 
is presented.  
 Section \ref{sec:String theory amplitudes} is devoted to
 string theory amplitudes on toroidal backgrounds.
The equivalence between  conformal anomaly cancellation conditions on the 
string vertex operators and the generalized harmonic gauge conditions on the DFT 
fields
is determined.
 We compute three point string scattering  amplitudes of massless and massive 
states and show the complete agreement with the expansions in DFT. The 
comparison 
involves a huge number of terms and so it is performed with the help of a 
computer 
(cadabra  program \cite{Peeters:2006kp}).
The simple example of circle compactification is worked out explicitly and 
the manifestly T-duality invariant  effective action is also presented.
A discussion on the limitations and possible extensions of this work
and a brief outlook are contained in the
 concluding remarks in Section \ref{sec:Conclusions and Outlook}.

\section{Double Field Theory basics}
\label{sec:Double Field Theory basics}

In this section we briefly review some of the basic features of DFT that are 
needed in our 
discussion.

The theory is defined on a double space with coordinates ${\mathbb X}^{\cal 
M}=(\tilde 
x_{\hat\mu}, x^{\hat\mu})$, defined in the fundamental
representation of $O(D,D)$. Here ${\cal M}=0, \cdots, 2D-1$ and $\hat\mu=0, 
\cdots, 
D-1$.

The generalized tensors transform under generalized diffeomorphisms as
\small{\bea 
{\cal L}_V W^{{\cal M}\cdots {\cal N}} =V^{\cal P}\partial_{\cal P} W^{{\cal 
M}\cdots{\cal N}} + \
(\partial^{\cal M}V_{\cal P}-\partial_{\cal P}V^{\cal M})W^{{\cal P}\cdots{\cal 
N}}+ \cdots + \
(\partial^{\cal N}V_{\cal P}-\partial_{\cal P}V^{\cal N})W^{{\cal M}\cdots{\cal 
P}}.\nn\\
\label{gendif}
\eea}
The natural $SO(D,D)$  metric  
\begin{equation}
\eta_{\cal MN}=\begin{pmatrix}0 & \mathbb{I}\\
\mathbb{I} & 0\end{pmatrix}
\label{oddmetric}
\end{equation}
is invariant under the above generalized transformations. It 
can be decomposed into a positive-definite and a negative-definite metric, 
$\left.\eta\right|_{C_\pm}$, acting on each of the two $D$-dimensional 
orthogonal subspaces of the doubled space $E=C_+\oplus C_-$, that are generated 
by the coordinates 
${\mathbb X}^{\cal M}_\pm = x^{\hat\mu} \pm \tilde x_{\hat\mu}$. Making use of 
$\left.\eta\right|_{C_\pm}$, a positive-definite metric 
can be defined on $E$
\begin{equation}
\mathcal{H}_{\cal MN}\, =\, 
\left(\left.\eta\right|_{C_+}-\left.\eta\right|_{C_-}\right)_{\cal MN}\, =\, 
\begin{pmatrix}g^{-1} & -g^{-1}\, B\\
B\, g^{-1}& g-B\, g^{-1}\,B\end{pmatrix}\, ,\label{gm}
\end{equation}
with
\begin{equation}
\mathcal{H}_{\cal MP}\,\eta^{\cal PQ}\,\mathcal{H}_{\cal QN}\, =\, \eta_{\cal 
MN} \, .
\end{equation}
Under $O(D,D)$ transformations $h_{\cal M}{}^{\cal P}$, 
${\mathbb X} \rightarrow {h\mathbb X}$ and the fields change as
\begin{equation}
 \mathcal{H}_{\cal MN}({\mathbb X})\rightarrow h_{\cal M}{}^{\cal P}h_{\cal N}{}^{\cal Q} 
 \mathcal{H}_{\cal PQ}({h\mathbb X}),\qquad d({\mathbb X})\rightarrow d({h\mathbb X})
 \label{oddtransf}
\end{equation}

Upper and lower indices are lowered and raised with 
$\eta_{\cal MN}$ and its inverse $\eta^{\cal MN}$, respectively.

It is sometimes useful to express the metric 
$\mathcal{H}_{\cal MN}$ in terms of a vielbein
\begin{equation}
\mathcal{H}_{\cal MN}=E^{ \cal A}{}_{\cal M}\, S_{\cal  A B}\, E^{\cal  
B}{}_{\cal N}\ , \qquad 
E^{ \cal A}{}_{\cal M}=\begin{pmatrix}e & e\, B\\
0 & e^{-1}\end{pmatrix}\, ,\label{metric}
\end{equation}
where $g_{\hat\mu\hat\nu}=e^{ a}{}_{\hat\mu}\, s_{ a b}\, e^{ b}{}_{\hat\nu}$,
\begin{equation}
S_{ \cal A B}=\begin{pmatrix}s^{ab} & 0\\
0 & s_{ab}\end{pmatrix}
\end{equation}
and $s_{ a b}$ is the $D$ dimensional Minkowski metric. ${\cal A, B}, \cdots$ 
indices are lowered and 
raised with the flat $SO(D,D)$ metric defined as
\begin{equation}
\eta_{\cal A B}=E_{ \cal A}{}^{\cal M}\, \eta_{\cal MN}\, E_{\cal  B}{}^{\cal 
N}\, 
\end{equation}
and its inverse, respectively, which numerically coincide with (\ref{oddmetric}).

Since the Minkowski metric is invariant under Lorentz $O(1,D-1)$ transformations, the 
metric $S_{ \cal A B}$ is invariant under double transformations $O(1,D-1)\times O(D-1,1)$
and as a result the  generalized metric $\mathcal{H}$ parametrizes the coset 
\begin{equation}
\frac{O(D,D)}{O(1,D-1)\times O(D-1,1)}\, . \label{coset}
\end{equation}

From the transformation law (\ref{gendif}), the generalized metric transforms as
\bea 
{\cal L}_V \mathcal{H}_{\cal MN} =V^{\cal P}\partial_{\cal P} {\mathcal 
H}_{\cal MN} + \
(\partial_{\cal M} {V}^{\cal P}-\partial^{\cal P} V_{\cal M})\mathcal{H}_{\cal 
PN} + \
(\partial_{\cal N} {V}^{\cal P}-\partial^{\cal P} V_{\cal N})\mathcal{H}_{\cal 
MP} .
\label{gendifh}
\eea

In terms of $\mathcal{H}_{\cal MN}$, and 
keeping up to two derivatives,  the action 
of DFT in the $2D$-dimensional space $E$ can be expressed as\footnote{In 
the original frame formulation of DFT by Siegel 
\cite{Siegel:1993xq,Siegel:1993th} the action includes extra terms that are not 
contained 
in (\ref{action}). Up to total derivatives those can be recast as 
\cite{Geissbuhler:2013uka}
\begin{equation*}
\Delta S\, =\, \frac1{G_{DFT}}\int d^{D}xd^D\tilde x\, e^{-2d}\, 
\left[\frac12 (S_{\cal A\cal
B}-\eta_{\cal A\cal B})\eta^{\cal PQ}\, \partial_{\cal M} E^{\cal A}{}_{\cal 
P}\partial^{\cal M} E^{\cal 
B}{}_{\cal Q}\, + \, 4\partial_{\cal M} d\partial^{\cal M} d \, - \, 
4\partial_{\cal M}\partial^{\cal M} d\right]\, .
\end{equation*}
In the Appendix we show that these terms vanish once the level matching 
condition described below, is imposed, and therefore we do not consider them in 
this work.}
\begin{equation}
S=\frac1{G_{DFT}}\int d^{D}xd^{D}\tilde x\, e^{-2d}\,\mathcal{R}({\cal H},d)\, 
,\label{action}
\end{equation}
where the generalized Ricci scalar is given by \cite{Hohm:2010pp}

\begin{multline}
\mathcal{R} \, =\, 4\mathcal{H}^{\cal MN}\partial_{\cal M}\partial_{\cal N} d\, 
-\, 
\partial_{\cal M}\partial_{\cal N}\mathcal{H}^{\cal MN}\, -\, 
4\mathcal{H}^{\cal MN}\partial_{\cal M} 
d\partial_{\cal N} d\, +\, 4\partial_{\cal M}\mathcal{H}^{\cal 
MN}\partial_{\cal N} d\\
+\, \frac18 \mathcal{H}^{\cal MN}\partial_{\cal M} 
\mathcal{H}^{\cal KL}\partial_{\cal N}\mathcal{H}_{\cal KL}
\,-\, \frac12 
\mathcal{H}^{\cal MN}\partial_{\cal M}\mathcal{H}^{\cal KL}\partial_{\cal 
K}\mathcal{H}_{\cal NL} \, ,
\end{multline}
the generalized dilaton 
\begin{equation}
e^{-2d}=e^{-2\phi}\, \sqrt{-g}
\label{gdilaton}
\end{equation}
is an $O(D,D)$ scalar and $G_{DFT}$ will be defined below
\footnote{The overall constant $G_{DFT}$ was introduced in \cite{blair}.}.

Comparison with string 
theory, as we re-discuss in more 
detail below, requires the \emph{level matching condition} (LMC) 
\beq \label{wc}
\partial_{\cal M}\partial^{\cal M}\cdots = (N-\bar{N})\cdots\, ,
\eeq
where  $N$ and $\bar{N}$ are the left and right oscillator numbers of the 
string and the dots stand for fields or gauge parameters.
Given that $g,B$ and $\phi$ correspond to $N=\bar{N}=1$, we would 
expect  $N-\bar{N}=0$. However,   in a compact space this 
difference could be a non-vanishing integer. Even though this is a key 
ingredient of
symmetry enhancing at certain compactification radii (see \cite{aimnr}),  we 
will only consider  states satisfying $N-\bar{N}=0$
in the main body of this article. 
Introducing the $2D$-dimensional momentum vector 
$\mathbb{P}^{\cal M}=(\tilde p^{\hat\mu}, p_{\hat\mu})$, generated by the partial 
derivatives 
$-i(\tilde{\partial}^{\hat\mu}, \partial_{\hat\mu})$ acting on the corresponding field, 
the constraint reads 
\begin{equation}
||\mathbb{P}||^2\, \equiv\, \mathbb{P}^{\cal M} \mathbb{P}_{\cal M}\, = \, 
0\label{lmc}
\end{equation}
for $\mathcal{H}_{\cal MN}$ and $d$.  In general,
this constraint is not sufficient to ensure consistency. For 
instance, the product of fields generically does not  satisfy it and the 
generalized transformations 
(\ref{gendif}) fail to close.

This  failure can be expected from string theory.  Namely, many other terms 
(actually infinite) are expected to complete the effective action, containing 
higher derivatives but also higher spin fields. Hopefully, in the full 
action variations could compensate among different terms and the algebra would 
close.
But in the truncated theory involving only  massless fields with $N=\bar{N}=1$  
in the
non compact case, consistency constraints are necessary. One solution of these 
constraints is  the so-called section
condition
\be \label{sc}
\partial_{\cal M}\cdots \partial^{\cal M}\cdots =0
\, ,
\ee
where the dots stand for products of fields or gauge parameters.
  It implies that half of the coordinates drop from the theory. These 
coordinates can be chosen to be  the dual coordinates $\tilde x_{\hat\mu}$. This 
choice is named \textit{gravity frame}
since in this case the action (\ref{action}) simply reduces to 
eq.~(\ref{eh}) when ${\cal H}_{\cal MN}$ is parametrized as in
(\ref{gm}) and $G_{DFT}\equiv 2\kappa^2\int d^D\tilde x$.  

The section condition is 
sufficient to satisfy the closure constraints, but there are more general 
solutions \cite{gm,{Geissbuhler:2013uka}}  when there is a compact 
sector. 
It is important to stress that (\ref{action}) describes more physical 
degrees of freedom than the 
standard $D$-dimensional action (\ref{eh}) for $g$, $B$ and $\phi$. 
Indeed, by introducing  coordinates $\tilde x_{\hat\mu}$, and their corresponding 
partial derivatives $\tilde{\partial}^{\hat\mu}$, fields can carry momentum along 
these directions 
 and the backgrounds can also depend on 
these coordinates. Such dependence is not an artifact of DFT: backgrounds 
with non-trivial dependence on the coordinates $\tilde{x}^{\hat\mu}$ cannot be 
described in terms of $D$-dimensional gravity, but are however expected to be 
consistent solutions of string theory. In particular, such backgrounds lead 
upon 
compactification to fully consistent effective gauged gravities with momenta 
along the internal coordinates associated  to winding 
excitations. DFT contains more degrees of freedom  than $D$-dimensional 
gravity, and in 
particular, it allows to compute observables and describe settings that 
cannot be accounted for in standard $D$-dimensional theories.

In what follows we will compactify DFT on generic tori with 
constant background fields and fluctuations around them. The constraints to be 
used will be extracted from comparison with string theory results.
In our computations,  the section condition must be imposed in the spacetime 
sector but only the LMC 
constraint is required in the toroidal compact space. This appears to be 
consistent if fluctuations are considered only up to third order. 
When going to higher orders, the 
 failure  of the gauge algebra to close should be interpreted as an indication 
that new degrees of 
freedom must be included. A brief discussion on this issue is offered in the
concluding 
remarks.

\subsection{ Einstein frame and harmonic coordinates}

The generalized metric $\mathcal{H}_{\cal MN}$ defined in (\ref{gm}) contains 
the 
$g$ and $B$ fields, and the generalized dilaton $d$  involves $\phi$. We can 
combine 
both $d$ and $\mathcal{H}_{\cal MN}$ into a single generalized Einstein-frame 
metric 
$\tilde{\mathcal{H}}_{\cal MN}$ with non-zero determinant. For that aim, we 
perform 
a Weyl transformation
\begin{equation}
\tilde{\mathcal{H}}_{\cal MN} = e^{2\Omega}\, \mathcal{H}_{\cal MN}\, ,
\end{equation}
under which a tensor with 
conformal weight $\Delta_\Phi$  transforms as
\begin{equation}
\tilde{\Phi} = e^{-\Omega\Delta_{\Phi}}\, \Phi \, .
\end{equation}
We list the conformal weights of the tensors  introduced in the previous 
section  in table \ref{tabla1}. 
\begin{table}[!ht]
\begin{center}
\begin{tabular}{|c||c|c|c|c|c|}
\hline
$\Phi$ & $\mathcal{H}_{\cal KL}$ & $\eta_{\cal KL}$ & $E^{\cal A}{}_{\cal M}$ & 
$S_{\cal A\cal 
B}$ & $\eta_{\cal A\cal B}$\\
\hline
$\Delta_\Phi$ & $-2$ & $-2$ & $-1$ & $0$ & $0$\\
\hline
\end{tabular}
\caption{Conformal weights of the various tensors that appear in DFT. 
\label{tabla1}}
\end{center}
\end{table}

Making use of these transformations, one can easily check that 

\begin{multline}
\mathcal{R}=e^{2\Omega}\left[\tilde{\mathcal{R}}\, -\, 2\partial_{\cal M} 
\tilde{\mathcal{H}}^{\cal MN}\partial_{\cal N}\Omega\, -\, 
2\tilde{\mathcal{H}}^{\cal MN}\partial_{\cal M}\partial_{\cal N}\Omega\, -\, 
(2+D)\tilde{\mathcal{H}}^{\cal MN}\partial_{\cal M}\Omega \partial_{\cal 
N}\Omega\right.\\
\left.+\, 8\tilde{\mathcal{H}}^{\cal MN}\partial_{\cal M}\Omega\partial_{\cal 
N} d\, -\, \frac12 
\tilde{\mathcal{H}}_{\cal KL}\tilde{\mathcal{H}}^{\cal MN}\partial_{\cal 
M}\tilde{\mathcal{H}}^{
\cal KL}\partial_{\cal N}\Omega\right]\, .
\end{multline}

Taking $\Omega=d$ and integrating by parts,
we can express (\ref{action}) in the Einstein frame as
\begin{equation}
S=\frac1{G_{DFT}}\int d^{D}xd^D\tilde x\,
\hat{\mathcal{R}}(\tilde {\cal H}, d)\, ,\label{actione}
\end{equation}
where
\bea
\hat{\mathcal{R}}(\tilde {\cal H}, d)& =& (2-D)\tilde{\mathcal{H}}^{\cal 
MN}\partial_{\cal M} 
d\partial_{\cal N} d\, -\, \frac12 
\tilde{\mathcal{H}}_{\cal KL}\tilde{\mathcal{H}}^{\cal MN}\partial_{\cal 
M}\tilde{\mathcal{H}}^{\cal
KL}\partial_{\cal N} d+\partial_M\partial_N{\tilde {\cal H}^{MN}}\nn\\
&& + \, \frac18 \tilde{\mathcal{H}}^{\cal MN}\partial_{\cal M} 
\tilde{\mathcal{H}}^{\cal KL}\partial_{\cal N}\tilde{\mathcal{H}}_{\cal KL}
\,-\, \frac12 
\tilde{\mathcal{H}}^{\cal MN}\partial_{\cal M}\tilde{\mathcal{H}}^{\cal 
KL}\partial_{\cal K}\tilde{
\mathcal{H}}_{\cal NL}\, .
\eea

This action (\ref{actione}) behaves similarly to the more 
familiar Einstein-Hilbert action in many aspects. In particular, the  equations 
of motion are 
greatly simplified by taking a harmonic coordinate condition to fix the gauge 
freedom under generalized diffeomorphisms. This can be achieved by requiring 
the 
coordinates ${\mathbb X}^{\cal R}$ to be solutions of the Laplacian equation
\begin{equation}
\partial_{\cal M}\left(\tilde{\mathcal{H}}^{\cal MN}\partial_{\cal 
N}\right){\mathbb X}^{\cal R}=0 \quad \Rightarrow 
\quad \partial_{\cal M}\tilde{\mathcal{H}}^{\cal MN}=0\, ,
\end{equation} 
which amounts 
to the gauge fixing condition\footnote{It can be shown that, in terms of the 
generalized connection of \cite{Siegel:1993xq,Siegel:1993th,Hohm:2011si}, this 
is 
equivalent to requiring $$\mathcal{H}^{\cal MP}\Gamma_{\cal MP}{}^{\cal Q}=0 
.$$}
\begin{equation}
\partial_{\cal M}\mathcal{H}^{\cal MN}-2\mathcal{H}^{\cal MN}\partial_{\cal M} 
d=0\, ,\label{harmonic}
\end{equation} 
when written in terms of the metric $\mathcal{H}_{\cal MN}$ and the scalar $d$.
Alternatively this equation can be expressed as
\begin{equation}
\partial_{\cal M} d\, =\, \frac12\mathcal{H}_{\cal MN}\partial_{\cal 
R}\mathcal{H}^{\cal NR}\, =\, 
-\frac12 \mathcal{H}^{\cal NR}\partial_{\cal R}\mathcal{H}_{\cal MN}\, .
\end{equation}
Making use of these conditions and integrating by parts, the action 
(\ref{action}) can be expressed in harmonic coordinates in a particularly 
compact form
\begin{equation}
S\, =\, \frac1{G_{DFT}}\int d^{D}xd^D\tilde x\, e^{-2d}\left[\frac18 
\mathcal{H}^{\cal MN}\partial_{\cal M}\mathcal{H}^{\cal KL}\partial_{\cal 
N}\mathcal{H}_{\cal KL}\, - \, 
\frac12 
\mathcal{H}^{\cal MN}\partial_{\cal M}\mathcal{H}^{\cal KL}\partial_{\cal 
K}\mathcal{H}_{\cal NL}\right]\, ,
\label{t1}
\end{equation}
or, in Einstein-like frame,
\bea
S_{DFT}&=&\frac1{G_{DFT}}
\int d^{D}xd^D\tilde x \left(
\frac18\tilde{\cal H}^{\cal MN}\partial_{\cal M}\tilde{\cal H}^{\cal 
KL}\partial_{\cal N}\tilde{\cal 
H}_{\cal KL} 
-\frac12\tilde{\cal H}^{\cal MN}\partial_{\cal M}\tilde{\cal H}^{\cal 
KL}\partial_{\cal K}\tilde{\cal 
H}_{\cal LN}
\right.\nn\\
&&~~~~~~~~~~~~~~~~~~~~~~\left.+~(2-D)\tilde{\cal H}^{\cal MN}\partial_{\cal M} d 
\partial_{\cal N} d\right)\, .
\label{dfthgef}
\eea

It is also interesting to express the gauge fixing condition in 
terms of $g$, $B$ and $\phi$. For standard $D$-dimensional gravity backgrounds 
with $\tilde p^{\hat\mu}=0$, one may easily check that its components reduce to
\begin{align}
\partial_{\hat\nu}\left(\sqrt{-g}\, g^{\hat\mu\hat\nu}e^{-2\phi}\right)&\, =\, 
0\, ,\label{gauge}\\
g^{\hat\mu\hat\nu}\partial_{\hat\nu} B_{\hat\lambda\hat\mu}&\, =\, 0\, .\nonumber
\end{align}
In particular, for vanishing dilaton $\phi = 0$, the first equation is 
 the usual  harmonic  gauge fixing condition of General Relativity. More 
generally, for generic DFT backgrounds, the gauge fixing conditions for $B$, 
$g$ and $\phi$ read
\begin{align}
\partial_{\hat\nu}\left(\sqrt{-g}\, g^{\hat\mu\hat\nu}e^{-2\phi}\right)&\, =\,  
\tilde{\partial}^{\hat\lambda}\left(\sqrt{-g}\, g^{\hat\mu\hat\sigma}
B_{\hat\sigma\hat\lambda}\, 
e^{-2\phi}\right)=0\, ,\\
\tilde{\partial}^\nu\left(\sqrt{-g}\, g_{\hat\mu\hat\nu} e^{-2\phi}\right) & \, = \, 
-e^{-2\phi} g^{\hat\sigma\hat\nu}\left(\partial_{\hat\nu} - 
B_{\hat\nu\hat\lambda}\tilde{\partial}^{\hat\lambda}\right)B_{\hat\mu\hat\sigma}=0 \, .\nonumber
\end{align}

\section{Perturbative DFT}
\label{sec:Perturbative DFT}

The physical content of a quantum field theory can be recast in terms of its 
S-matrix elements, that are usually computed perturbatively. In the particular 
case of General Relativity, perturbative computations are however specially 
complex due to the huge number of vertices, rendering most of the brute force 
computations of scattering amplitudes infeasible. Fortunately, the field theory 
limit of Kawai-Lewellen-Tye (KLT) relations \cite{Kawai:1985xq} allows to 
express gravity amplitudes in terms of two copies of gluon amplitudes, which 
are 
much simpler to compute. In particular, starting from gluon amplitudes and 
using 
KLT relations, it has been possible to construct a Lagrangian for gravity 
\cite{Bern:1999ji}. The resulting Lagrangian is particularly simple and is 
related to the usual Einstein-Hilbert action by non-linear redefinitions and 
gauge fixing similar to those used in \cite{vandeVen:1991gw}. Moreover graviton 
spacetime indices can be split into two types (left and right), in such a way 
that 
contractions do not mix indices of different type.

KLT relations originate from the fact that the integrand of a closed string 
amplitude involves two open string components, corresponding to left and 
right movers. It is then natural to expect that this hidden 
simplification of gravity amplitudes  also holds in DFT. Indeed, this is 
already manifest in the extreme simplicity of the Lagrangian (\ref{t1}).  
To be more specific, let us 
split $\mathcal{H}_{\cal MN}$ into background $\overline{\mathcal{H}}_{\cal 
MN}$ and 
quantum fluctuations $\hat h_{\cal MN}$,
\begin{eqnarray}
 \mathcal{H}_{\cal MN}\, = \, \overline{\mathcal{H}}_{\cal MN} \, + \, \hat 
h_{\cal MN}\,, \qquad
 d=\overline{d}+\hat d\, . \label{fluctexp}
\end{eqnarray}

For simplicity, we consider $\overline{\mathcal{H}}_{\cal MN}$ and 
$\overline{d}$ to be constant.
Due to the presence of  two metrics, namely $\mathcal{H}_{\cal MN}$ 
and $\eta_{\cal MN}$,
(\ref{fluctexp}) can be
inverted in two different ways: by making use of $\eta^{\cal 
MN}$ 
or by using the geometric series for matrices. We thus obtain for the inverse
\begin{equation}
\mathcal{H}^{\cal MN}\, = \,  \overline{\mathcal{H}}^{\cal MN} \, + \, \hat 
h^{\cal MN}\, = \, 
\overline{\mathcal{H}}^{\cal MN} - \, \hat h^{\dot {\cal M}\dot {\cal N}}  +\, 
\hat  h^{\dot{\cal  M}}{}_{\cal P}\, \hat h^{\dot 
{\cal P}\dot {\cal N}} - \, \hat h^{\dot {\cal M}}{}_{\cal Q}\,\hat  h^{\dot{ 
\cal Q}\dot{\cal  P}}\,\hat  h_{\cal P}{}^{\dot{\cal N}} + \, \ldots \, ,
\end{equation}
where we have introduced the short-hand notation $
A^{\dot{\cal M}}\, \equiv\, \overline{\mathcal{H}}^{\cal MN}A_{\cal N} \ ,   \, A_{\dot{\cal M}}\, 
\equiv\, \overline{\mathcal{H}}_{\cal MN}A^{\cal N}$,
and it is useful to note that, up to first order,
\begin{equation}
\hat h^{\cal MN}=\eta^{\cal MP}\eta^{\cal NQ}\hat h_{\cal QP}= 
-\overline{\mathcal{H}}^{\cal MP}\overline{\mathcal{H}}^{\cal NQ}\hat h_{\cal 
QP}=-\hat 
h^{\dot{\cal M} \dot {\cal N}}\, .
\label{etaeme}
\end{equation} 
The single field $\hat h^{\cal MN} \equiv \eta^{\cal MP}\eta^{\cal MQ} \hat 
h_{\cal PQ}$ therefore encodes an 
infinite set of operators when expressed in terms of the background  metric 
$\overline{\mathcal{H}}_{\cal MN}$.

Note also that by construction 
$\overline{\mathcal{H}}^{\cal MN}\overline{\mathcal{H}}_{\cal NQ}=\delta^{\cal 
M}{}_{\cal Q}$, however 
$\hat h^{\cal MN}\hat h_{\cal NQ}\neq \delta^{\cal M}{}_{\cal Q}$. Instead, one 
may easily check the following 
relation
\begin{equation}
\hat h^{\cal MN}\hat h_{\cal NQ} \, = \, -\left(h^{\dot{\cal M}}{}_{\cal Q}\, 
+\, h^{\cal M}{}_{\dot {\cal Q}}\right) \, .
\end{equation}

 For comparison with string theory results, it  proves convenient to look at 
fluctuations in the so-called modified Einstein frame, namely the  Einstein frame
 discussed above with the vacuum value of the generalized dilaton 
 $e^{-2\overline d}$  extracted out \footnote{In what follows Einstein frame 
  means modified Einstein frame.}.
Thus,  the generalized metric is, up to first order
\begin{equation}
 \tilde{\cal H}_{\cal MN}= {\bar{\tilde{\cal H}}}_{\cal MN}+ \hat{\tilde{ 
h}}_{\cal MN}=
 {\bar {\cal H}}_{\cal MN} +({\hat  h}_{\cal MN}+2\hat d{\bar 
{\cal H}}_{\cal MN} )\, .\label{fluctexpef}
\end{equation}

\subsection{ Expansion of  DFT in  fluctuations}
Following the discussion above, by using (\ref{fluctexpef}) we expand the DFT 
harmonic gauge fixed action (\ref{dfthgef}) into background and quantum 
fluctuations. We get,
up to third order in fluctuations,
\bea
L_{DFT}
&=& 
\frac18{ \bar{\tilde{\cal H}}}^{{\cal M}{\cal N}}\partial_{\cal {M}}\hat{\tilde h}^{\cal K L}
\partial_{ \cal N}\hat {\tilde h}_{\cal KL} -(D-2){\bar{\tilde{\cal H}}}^{{\cal M}{\cal N}}
\partial_{{\cal M}}\hat d \partial_{ \cal N}\hat d\nn
\\
&& -\frac12 \hat{\tilde  h}^{{\cal M} {\cal  N}}\partial_{ \cal 
M}\hat {\tilde h}^{{\cal K}  {\cal L}}\partial_{ 
\cal K}\hat
{\tilde h}_{\cal NL}+ \frac18  \hat{\tilde h}^{{\cal M} 
{\cal N}}
 \partial_{\cal M } \hat {\tilde h}^{ {\cal K} {\cal L}} \partial_{\cal 
N} \hat{\tilde h}_{\cal KL} -(D-2)\hat {\tilde h}^{ \cal M N}
\partial_{ \cal M}\hat d \partial_{ 
\cal M}\hat d\, . \nn\\
\label{ldft1}
\eea

Recall that in terms of fields, the fluctuations
$\hat h_{\cal MN}=\hat h_{(1)\cal MN}+\hat h_{(2)\cal MN}+\dots$
contain contributions from higher orders.
In particular,  terms quadratic in $\hat h_{\cal MN}$  could give third order 
interaction terms. However, 
this is not the case. Actually, integrating by parts the term
$
\frac14 
 \overline{\tilde{\mathcal{H}}}^{\cal MN}
\partial_{\cal M} \hat {\tilde h}_{(2)}^{\cal KL}\partial_{\cal N} \hat {\tilde h}_{(1)\cal KL}$,
one gets
 the equations of motion (see (\ref{eom}) below), and so 
this cubic term vanishes on shell. The same conclusion holds for the second 
term.
Therefore, the third order terms in the action  involve only the first order  
fluctuations of the generalized fields, and we finally have the Lagrangian (\ref{ldft1})
with  $\hat {\tilde h}_{\cal KL}=\hat {\tilde h}_{(1)\cal KL}$.

Before compactification,  in a flat background
\be
 \overline{\mathcal{H}}_{\cal M \cal N} = \left(\begin{matrix}  
\eta^{\hat\mu\hat\nu} & 0 
 \\
0 &  \eta_{\hat\mu\hat\nu}\end{matrix}\right) \, ,
\label{GenMetEffectivebc}
\ee
and to first order in fluctuations, 
$g_{\hat\mu\hat\nu}=\eta_{\hat\mu\hat\nu}+h_{\hat\mu\hat\nu},
B_{\hat\mu\hat\nu}= b_{\hat\mu\hat\nu}$, we have
\be
 { \hat h}_{\cal M \cal N} = \left(\begin{matrix}  { \hat h}^{\hat\mu\hat\nu} & 
 { \hat h}^{\hat\mu } _{\ \ \hat\nu} & \\
{ \hat h}_{\hat\mu}{}^{\hat\nu} &  \hat{ h}_{\hat\mu \hat\nu}  & 
\end{matrix}\right) 
= \left(\begin{matrix}  -h^{\hat\mu \hat\nu} & -
 \eta^{\hat\mu \hat\rho} b_{\hat\rho\hat\nu} &  \\
-  \eta^{\hat\nu\hat\rho} b_{\hat\rho\hat\mu} &  h_{\hat\mu\hat\nu} \end{matrix}\right)
\, .\label{GenPertMetEffective1}
\ee

Then
from the second order terms in fluctuations
 and imposing the strong constraint in the gravity 
frame (namely, dropping
the dependence on the $\tilde x_{\hat\mu}$ coordinates), we recover the  quadratic 
terms in the
action (\ref{eh}) in the de
Donder gauge \cite{harmonicg,ortin}. Actually,  in the string frame we get
\begin{equation}
S=\frac{1}{2\kappa^2}\int d^D x\,  e^{-2{\bar \phi}}
\left[\partial_{\hat\sigma}\left(\frac{h_{\hat\nu}{}^{\hat\nu}}{2}-2\hat\phi\right)\partial^{
    \hat\sigma}\left(\frac{h_{\hat\rho}{}^{\hat\rho}}{2}-2\hat\phi\right)
\, -\, \frac12 
\partial_{\hat\sigma} 
( h_{\hat\nu\hat\lambda}+b_{\hat\nu\hat\lambda})\partial^{\hat\sigma}
( h^{\hat\nu\hat\lambda}+b^{\hat\nu\hat\lambda}
)\right]\, .
\end{equation}
Transforming this action into momentum space, we  obtain the propagators 
for $h$, $b$ and $\hat \phi$
\begin{eqnarray}
 D^{h}_{\hat\mu\hat\nu; \hat\rho \hat\sigma}&= &
\frac{e^{-2{\bar \phi}}}{4}\frac{\eta_{\hat\mu\hat\rho}
\eta_{\hat\nu\hat\sigma}}{p^2}\, ,\nn\\
 D^{b}_{\hat\mu\hat\nu; \hat\rho \hat\sigma}&=  &
\frac{e^{-2{\bar \phi}}}{4}\frac{\eta_{\hat\mu\hat\rho}
\eta_{\hat\nu\hat\sigma}-\eta_{\hat\mu\hat\sigma}\eta_
{\hat\nu\hat\rho}}{p^2}\, ,\nonumber\\
D^{2\hat\phi-\frac{h^{\hat\nu}{}_{\hat\nu}}{2}}&=&\frac{4e^{-2{\bar \phi}}}{p^2}\, .\nn
\end{eqnarray}
The first lesson to be drawn from this calculation is that the strong 
constraint must be imposed on the
space-time coordinates in order to recover ordinary gravity theories, as 
expected.

\subsection{Generalized Kaluza-Klein compactification}

The generalized Kaluza-Klein (GKK) decomposition of the generalized metric 
reads
\be
 {\cal H}_{\cal MN} = \left(\begin{matrix}  g^{\mu \nu} & -
 g^{\mu \rho} c_{\rho \nu} & -  g^{\mu \rho}
 A_{N\rho} \\
-  g^{\nu\rho} c_{\rho \mu} &  g_{\mu \nu} +
 A^N{}_\mu  {\cal M}_{NP}  A^P{}_\nu  + 
c_{\rho\mu}  g^{\rho \sigma}  c_{\sigma \nu} & 
{\cal M}_{NP}  A^P{}_\mu +  A_{N\rho} 
g^{\rho\sigma}  c_{\sigma \mu} \\ -  g^{\nu \rho} 
A_{M\rho} &  {\cal M} _{MP}  A^P{}_{\nu} +  A_{M
\rho} g^{\rho \sigma}  c_{\sigma \nu}&  {\cal
M}_{MN} +  A_{M \rho}  g^{\rho \sigma}
A_{N\sigma}\end{matrix}\right) \, ,\label{GenMetEffectivekk}
\ee
where now the ${\cal M, N}$  indices split into spacetime $\mu, \nu, \cdots $ 
indices taking the values $0, \cdots, 
d-1$, and  internal doubled indices $M, N, \cdots = 1, \cdots , 2(D-d)$. We have 
introduced the combination $ c_{\mu\nu} = 
b_{\mu \nu} + \frac{1}{2}  A^N_\mu  A_{N\nu}$,
$ A^N_\mu$ denote the vectors and  $
{\cal M}_{MN}$ is the  scalar matrix defined below.

In terms of components, the  constant generalized background metric reads now
\be
 \overline{\cal{H}}_{\cal M \cal N} = \left(\begin{matrix}  \eta^{\mu \nu} & 
0 & 0
 \\
0 &  \eta_{\mu \nu} 
& 
0\\ 0&  0&
\overline{\cal M}_{MN}\end{matrix}\right) \, ,\label{GenMetEffective}
\ee
with 
\be
 \overline{\cal M}_{MN} \  \ = \
  \begin{pmatrix} G^{mn}  
    &-G^{mp}
    B_{pn}\\[0.5ex]
    B_{mp}  G^{pn}  
  & G_{mn}-B_{mp}
    G^{pq}  B_{qn}\end{pmatrix}\, ,\label{Meffective} 
\ee
where $m, n,\cdots =1, \cdots, D-d$.
The fluctuations
up to first order are
\be
 { \hat h}_{(1)\cal M \cal N} = \left(\begin{matrix}  { \hat h}_{(1)}^{\mu \nu} 
& 
 { \hat h}^{\mu } _{(1) \nu} & { \hat h}^{\mu}_{(1) N} \\
{ \hat h}_{(1)\mu}^{\ \ \nu} &  { \hat h}_{(1)\mu \nu}  & 
{ \hat h}_{(1)\mu  N} \\ { \hat h}_{(1)M}^{\ \ \nu } &  { \hat h}_{(1)M\nu } & 
{ 
\hat h}_{(1) MN} \end{matrix}\right) 
= \left(\begin{matrix}  -h^{\mu \nu} & -
 \eta^{\mu \rho} b_{\rho \nu} & -  \eta^{\mu \rho}
 A_{N\rho} \\
-  \eta^{\nu\rho} b_{\rho \mu} &  h_{\mu \nu}  & 
\overline{ \cal M}_{NP}  A^P{}_\mu\\ -  \eta^{\nu \rho} 
A_{M \rho} &  \overline{\cal M} _{MP}  A^P{}_{\nu} &  {
h}_{MN} \end{matrix}\right)
\, .\label{GenPertMetEffective}
\ee
 The matrix ${
h}_{MN}$ encoding the scalar field content reads
 \begin{equation}
h_{MN}=
\begin{pmatrix}
-G^{n k}h_{k l}G^{l m} & -G^{nk}b_{km} + G^{n k}h_{k s}G^{s l}B_{l m}\\
-B_{n s}G^{s l}h_{l k}G^{k m} + b_{n l}G^{l m} & h_{n m} - b_{n l}G^{l k}B_{k m}  
+B_{n k}G^{k s}h_{s r}G^{r b}B_{b m} - B_{n k}G^{k l}b_{l m}
\end{pmatrix}
\label{scalarsmatrix}
\end{equation}
where   $h_{n l},b_{n l}$ are the scalar fields derived from the higher dimensional 
graviton and antisymmetric fields, respectively.

From the definition of the generalized dilaton (\ref{gdilaton}) and recalling that
$d={\overline d}+{\hat d}$, we have 
\bea
e^{-2\overline d}&=&e^{-2 \phi_0}{\sqrt{det {G}}}\, ,\label{cgdilaton}\\
{\hat d }&=& \hat {\phi} -\frac14 h^{\mu}{}_{\mu}\, .
\eea

In Einstein frame,
 the only fluctuations that are modified  are
\bea
\hat{\tilde{ h}}_{\mu\nu}&=&({ h}_{\mu\nu}+2{\hat d}\,\eta_{\mu\nu})\equiv 
\tilde h_{\mu\nu}\, ,\nn\\
\hat{\tilde{ h}}^{\mu\nu}&=&(-{ h}^{\mu\nu}+2{\hat d}\, 
\eta^{\mu\nu})\equiv -\tilde h^{\mu\nu}\, ,\nn\\
\hat{\tilde{ h}}_{MN}&=&({ h}_{MN}+2{\hat d} \,\overline{\cal M}_{MN})\equiv \tilde 
h_{MN}\, .
\label{fluctef}
\eea

The harmonic 
gauge conditions in the Einstein frame
($
\partial^{\cal M} \tilde{\mathcal{H}}_{\cal MN}=0$)
 become, in terms of fluctuations,
\begin{equation}
\partial^{\cal M}  \hat{\tilde{ h}}_{\cal MN}=\partial_{\mu}\hat{\tilde{ 
h}}^{\mu}{}_{\cal N}+\partial^{L}\hat{\tilde{ h}}_{L\cal N}=0\, ,
\label{hcef}
\end{equation}
where we have used the strong constraint in the spacetime sector.
Therefore, when specifying  values for the index ${\cal N}$, we have
\bea
\partial_{\mu}\hat{\tilde{ 
h}}^{\mu\nu}+\partial_{N}\hat{\tilde{ h}}^{{ N}\nu}=0&\rightarrow& \partial_{\mu}{\tilde{ 
h}}^{\mu\nu}+\partial_{N} A^{{ N}\nu}=0\, ,\\
\partial_{\mu}\hat{\tilde{ 
h}}^{\mu}{}_{\nu}+\partial_{N}\hat{\tilde{ h}}^{N}{}_{\nu}=0
&\rightarrow& \partial^{\mu}{b}_{\mu\nu}-i(\mathbb{P}{\overline M}A)_{\nu}=0\, ,\\
\partial_{\mu}\hat{\tilde{ 
h}}^{\mu  N}+\partial_{M}\hat{\tilde{ h}}^{ MN}=0  &\rightarrow&
\partial^{\mu}A_{\mu}^{N}-i(\mathbb{P}{\overline M}\tilde h{\overline M})^{N}=0
\, .
\label{hgfields}
\eea

We will discuss  the link between this set of equations and the 
vanishing of conformal anomalies in string theory in section 
(\ref{sec:String theory amplitudes})
below.

\section{Toroidal compactification}
\label{sec:Toroidal compactification}

We  consider the  mode expansion of fields on an internal $2n$-dimensional 
double torus with 
 constant background (metric, dilaton  and antisymmetric fields) turned on.
 It corresponds to a compactification on  $2(D-d)=2n$ circles, which are 
generically non-orthogonal
 since the background metric is in general non-diagonal\footnote{Here we 
 consider dimensionful internal coordinates whereas the metric is 
dimensionless. 
 Alternatively, we could absorb the dimensions in the metric just by redefining 
$\overline 
 G_{mn }\rightarrow \overline 
 G_{mn }R^{(m)} R^{(n)}$}.
 
The internal coordinates $\mathbb{Y}^M=(\tilde y_{m},y^{m})$ have periodicity 
\bea
\tilde y_m \sim \tilde y_m+2\pi \tilde R_{(m)}\, ,\qquad y^m\sim y^m+2\pi 
R^{(m)}\, ,\label{pc}
\eea
where $R^{(m)}$ and $\tilde R_{(m)}=\alpha ' {R^{(m)}}^{-1}$ denote the radii 
of the $m$-th cycle 
and its dual, respectively. The internal momenta are encoded in the 
$O(n,n)$ vector  $\mathbb{P}_M$  of components
\bea
\mathbb{P}_{M}\equiv (\mathbb{P}_m,\mathbb{P}_{n+m})=(p_{m},\tilde 
p^{m} )= \left(\frac{n_m}{R^{(m)}},\frac{w^m}{\tilde R_{(m)}}\right)\, ,
\label{onnmomenta}
\eea
$n_m$ and $ w^m$ being the integer momentum and winding  numbers.

 On the torus background, the non-trivial identifications (\ref{pc}) are only 
preserved by $O(n,n)$ transformations with integer-valued matrix entries. Thus,  
the $O(n,n, {\mathbb R})$ symmetry is broken to the discrete $O(n,n,{\mathbb 
Z})$ group.

The mode expansion of the generalized metric  would 
be ${\cal H}(x,\mathbb{Y})={ \bar{\cal H}}+\hat h(x,\mathbb{Y})$ with 
\bea
    {\hat h}(x,\mathbb{Y})= \sum_{\mathbb{P}}{}^{\prime}
   {\hat h}^{( 
\mathbb{P})}(x)e^{i \mathbb{P}_{M} \mathbb{Y}^{ M}}\, ,
 \label{mexp}
 \eea
where the dependence on the dual space time coordinates $\tilde x_\mu$ has been 
dropped. The expansion of the component fields is 
\bea
g_{\mu\nu}(x, {\mathbb Y})&=&\eta_{\mu\nu}+\sum_{\mathbb P}{}^\prime 
h^{({\mathbb P})}_{\mu\nu}(x)e^{i \mathbb{P}_{M} \mathbb{Y}^{ M}}\, ,\\
b_{\mu\nu}(x, \mathbb Y)&=&\sum_{\mathbb{P}}{}^{\prime}
   {b}^{( 
\mathbb{P})}_{\mu\nu}(x)e^{i \mathbb{P}_{M} \mathbb{Y}^{ M}}\, ,
\eea
and similarly for ${\hat d}(x, \mathbb{Y})$, gauge parameters, etc.

The sum over $\mathbb{P}$ involves, in principle,  all  integer 
values of momenta and windings
$(n_m,w^m)$. Possible constraints are  indicated with  a prime on the sum.
Also,  since all the fields we are dealing with are real,  we require
$
{ \cal H}^{( -\mathbb{P})}(x)={\cal H}^{( \mathbb{P})}(x)^*$. 

Remember that we have dropped the field dependence on dual space-time coordinates,
or in DFT words, we have imposed the strong constraint in order to stay in the 
gravity frame. This means that there will be a $\frac{1}{2\kappa^2}$ overall factor in 
the action, where  $\kappa$ is now the gravitational constant in $d+2n$ 
dimensions. 
In terms of the  DFT coupling above it would formally read 
$\frac{1}{2\kappa^2}=\frac{1}{G_{DFT}}\int d^d\tilde x$.

Due to the contributions from both, a  circle and its dual, 
the usual $R$ dependent volume factor of dimensional reduction is not present 
here,   and instead an $\ap$ factor is left, namely
\begin{equation}
 d^{2n}\mathbb{Y}=\Pi_{i=1}^n 
\frac{d y^i}{2\pi R_{i}}
\frac{d\tilde y_i}{2\pi \tilde R_{i}}=
\Pi_{i=1}^n\frac{1}{(2\pi)^2 \ap} {dy^i}{d\tilde y_i}\, .
\end{equation}
 Furthermore, we use that
\be
\int d^{2n}\mathbb{Y} e^{i 
(\mathbb{P}_{M}+\mathbb{Q}_{M}) \mathbb{Y}^{ 
M}}=\delta^{2n}(\mathbb{P}_{M}+\mathbb{Q}_{M})\, ,
\ee
since $\int_0^{2\pi R_i}\frac{dy_i}{2\pi R_{i}}=1$.
 We will see below that the dependence on radii 
shows up when vector fields are redefined in order to have integer  $U(1)$ 
charges.
 Also a scaling factor appears through the expectation value of the 
generalized dilaton $e^{-2\bar d}$ containing both the  determinant of the 
background metric $\bar G$ and  dilaton $\bar \phi$ fields.

\subsection{Quadratic terms and masses}

We first concentrate on the quadratic terms in the 
action. Inserting the GKK
expansion in the first line of the Lagrangian (\ref{ldft1}), we 
obtain
\bea
S_{DFT}^{(2)}
&=& \frac1{2\kappa^2_d}\sum_{\mathbb P}{}^\prime\int d^dx
\left[ \hat 
d(x)^{(\mathbb{P})}(\partial_{\mu}\partial^{\mu}-\mathbb{P}_M\overline{\cal 
M}^{MN}\mathbb{P}_N )\hat 
d(x)^{(-\mathbb{P})}\right.\nn\\&&~~~~~~~~~~~~~~~~~~~~~ -
\left. \frac18
\hat {\tilde h}^{(\mathbb{P})\cal KL}(x) 
(\partial_{\mu}\partial^{\mu}-\mathbb{P}_M\overline{\cal 
M}^{MN}\mathbb{P}_N )\hat {\tilde h}^{(-\mathbb{P})}_{\cal KL}(x)\right]\, ,
\label{eom}
\eea
where we have redefined $\hat d \rightarrow (D-2)^{1/2}\hat d$, and by using 
(\ref{cgdilaton}),
\begin{equation}
 \frac1{2\kappa^2_d}=\frac1{2\kappa^2}e^{-2\overline d}.
\end{equation}

The  equations of motion read
\begin{equation}
 (\partial_{\mu}\partial^{\mu}-\mathbb{P}_M\overline{\cal M}^{MN}\mathbb{P}_N )
 \hat {\tilde h}^{(\mathbb{P})}_{\cal KL}(x)=0,\qquad 
(\partial_{\mu}\partial^{\mu}-\mathbb{P}_M\overline{\cal M}^{MN}\mathbb{P}_N )
 {\hat d}^{(\mathbb{P})}(x)=0 .
\end{equation}

Interestingly enough, these expressions not only reproduce the propagators for 
the gravity multiplet\footnote{A careful discussion about physical degrees of 
freedom is 
presented in next section.} but they also contain the 
propagators for GKK states. In particular,  we can identify the mass squared of 
the GKK ${(\mathbb{P})}$ modes\footnote{Here the dot refers to contractions 
with the internal metric 
$\overline {\cal 
M}$.}
  as 
\begin{equation}
M^2= -k^2= \mathbb{P}_M\overline{\cal M}^{MN}\mathbb{P}_N \label{mass}= 
{\mathbb{P}}^{\dot M} \mathbb{P}_M\, .
\end{equation}

This is exactly the mass squared of string states on generic toroidal  
backgrounds for
$N+\bar N-2=0$. We expect this condition is satisfied since we started with 
$N=\bar 
N=1$. However,  the string states also satisfy the LMC, namely
\begin{equation}
 \frac12\mathbb{P}_M\mathbb{P}^M=N-\bar N=0\, .
\end{equation}
Therefore, it appears that  in order to recover the string theory results, we 
must 
consider the following constrained GKK expansion
\bea
 {\hat h}(x,\mathbb{Y})&=&\sum _{\mathbb{P}} {\hat h}^{( 
\mathbb{P})}(x)e^{i \mathbb{P}_{M} \mathbb{Y}^{ M}}
\delta({\mathbb{P}}^2)\, ,
\label{cmexp}
 \eea
and similarly for $\hat d(x,\mathbb{Y})$.

Let us look at the transformation  of the compactified 
action under the generalized diffemorphisms (\ref{gendif}). From the discussion above, 
we know that this variation should be proportional to terms that vanish if 
the strong constraint $\partial_{\cal P}\otimes\partial^{\cal P}=0$ is  imposed. 
Moreover, since the space-time part already satisfies it, the transformation 
must be proportional to  $\partial_{ P}\otimes\partial^{ P}=0$, where now $P$ labels the
 internal compact coordinates. Since the variation is proportional to the gauge parameter,
it can be written as $\partial_{ P} \xi_{M} J^{P M}$, with $J^{P M}$ a product 
of  generalized metric and  dilaton fields with a  $\partial^{ P}$ derivative acting 
on one of them. By mode expanding the generalized fields, these derivatives 
lead to a $\mathbb{Q}^i_{P}\mathbb{Q}^{j P}$ factor times a 
$\delta^{2n}(\sum_i \mathbb{Q}^i)$ requiring total momentum conservation.
If up to third order  terms in fluctuations are kept in the action, momentum 
conservation and level matching $ {\mathbb{Q}^i}^2=0$ for each field (including $\xi_{M}$)
leads to $\mathbb{Q}^i\cdot \mathbb{Q}^j=0$ and we conclude that the action,
up to this order, is invariant  under generalized diffeomorphisms.

\subsection{Physical degrees of freedom}

The mass formula (\ref{mass}) is generic and does not allow us to isolate 
physical states.
For instance, $\hat{\tilde h}^{(\mathbb{P}) \mu}_{M}(x)$ seems to denote 
$2(D-d)$ massive vector states. 
However, we know that some of these vectors must be absorbed by the gravitational
and two-form fields to become massive.
Actually, the harmonic gauge condition allows  to identify the physical 
degrees of freedom.  
In order to see this, first recall the expected physical fields in lower 
dimensions.

A symmetric massless  two-tensor in $D$ dimensions has $(D-2)(D-1)/2$ degrees of 
freedom\footnote{We count  here the degree of freedom of the trace, associated to 
the dilaton field. We discuss the splitting of traceless and trace parts below, in 
order to compare with string theory results.}.
With $n$ compact  dimensions,  we can write
\bea
\frac12(D-2)(D-1)&=&\frac12(D-n-2)(D-n-1)+n(D-n-2)+\frac12n(n+1)\quad{\rm or}\nn\\
\frac12(D-2)(D-1)&=&\frac12(D-n-1)(D-n)+(n-1)(D-n-1)+\frac12n(n-1)\nn
\eea
Starting with the metric in $D$ dimensions, decomposing the indices into  
$D-n$ 
spacetime 
 and $n$  internal indices, for massless states (corresponding to zero 
modes in the  KK expansion) we  would have 
$\frac12(D-n-2)(D-n-1)$ d.o.f. for  $g_{ \mu  \nu}$, $ n$ vectors  $g_{\mu m}$ 
leading to 
$ n(D-n-2)$ d.o.f. and
$\frac12n(n+1)$ scalars $g_{mn}$, consistent with the first equation.
On the other hand,  if the states are massive,  we must decompose them as in 
the second 
equation, corresponding to a massive symmetric two-tensor, $n-1$ massive vectors 
and $\frac12n(n-1)$ scalars.
We can understand this combination by interpreting that a scalar is eaten by a 
massless vector to become massive,  leaving $\frac12n(n+1)-n=\frac12n(n-1)$ 
scalars and  $n$ massive vectors with $(D-n-1)$ degrees of freedom. However, 
one 
of these vectors is eaten by the massless graviton to become massive, leaving a 
massive two-tensor with $\frac12(D-n-2)(D-n-1)+(D-n-1)=\frac12(D-n-1)(D-n)$ 
d.o.f., 
and $n-1$ massive vectors.

A similar computation can be done for the antisymmetric tensor.  Namely, a 
massless two-tensor with $\frac12 (D-2)(D-3)$ d.o.f. 
can be decomposed as
\small{\bea
\frac12(D-2)(D-3)&=&\frac12(D-n-2)(D-n-3)+n(D-n-2)+\frac12n(n-1)\,\\
\frac12(D-2)(D-3)&=&\frac12(D-n-1)(D-n-2)+(n-1)(D-n-1)+\frac12(n-2)(n-1). \nn
\eea}
The first equation leads to the familiar KK decomposition in terms of a 
massless  two-tensor $b_{ 
\mu  \nu}$, $n$ massless vectors
$b_{ \mu  m}$ and $\frac12n(n-1)$ massless scalars $b_{ mn}$. For the massive 
case,
a massless antisymmetric tensor eats a massless vector, leaving a massive 
antisymmetric tensor with 
$\frac12(D-n-2)(D-n-3)+(D-n-2)=\frac12(D-n-1)(D-n-2)$ d.o.f.
The $n-1$  massless vectors left eat $n-1$ scalars to become $n-1$ massive 
vectors, leaving
$\frac12n(n-1)-(n-1)=\frac12(n-2)(n-1)$ massive scalars.

On the whole, a  massive GKK level is characterized by the generalized momentum
$\mathbb{P}$, with $ \mathbb{P}^2=0$, and it contains a spin two symmetric 
tensor (which can be decomposed into a traceless tensor + trace), an antisymmetric tensor, $2(n-1)$ vectors and 
$n(n-1)$ scalars, all mass degenerate with mass $M^2= \mathbb{P}{\overline{\cal M} }\mathbb{P}$. 
Note that a non-equivalent level $ \mathbb{P}'=h\mathbb{P}$ will have the same mass
if $h$ is an 
$O(n,n)$ transformation, namely $h$ is a 
duality transformation.
Recall that, in the  $n=1$ double circle case no extra massive vectors or 
scalars are present. 

In (both spacetime and internal) momentum space,  the generalized harmonic gauge
conditions (\ref{hgfields}) for the modes  $\hat{\tilde{ 
h}}^{(\mathbb{P})}_{\cal MN}(k)$ 
read
\bea
k^{\mu}\hat{\tilde{ 
h}}^{(\mathbb{P})}_{\mu \cal N}(k)+ (\mathbb{P}\hat{\tilde{ 
h}}^{(\mathbb{P})})_{\cal N}(k)
=k^{\mu}[ \hat{\tilde{ 
h}}^{(\mathbb{P})}_{\mu \cal N}(k) -\frac{1}{M^2}k_{\mu}(\mathbb{P}\hat{\tilde{ 
h}}^{(\mathbb{P})})_{\cal N}(k)]=0\, ,\label{ae}
\eea
where we have used that $-k^2=M^2$ is the (squared) mass   of the states 
as given in (\ref{mass}).

This is an indication that there is a physical massive field 
\bea
\hat{ \tilde{ 
h}}^{\prime(\mathbb{P})}_{\mu \cal N}(k)=\hat{\tilde{ 
h}}^{(\mathbb{P})}_{\mu \cal N}(k) -\frac{1}{M^2}k_{\mu}(\mathbb{P}\hat{\tilde{ 
h}}^{(\mathbb{P})})_{\cal N}(k)+\dots \, ,\nn
\eea
(where $\dots$ indicate possible terms vanishing when contracted with 
$k^{\mu}$) or equivalently 
\bea
 \hat{\tilde{ 
h}}^{\prime(\mathbb{P})}_{\mu \cal N}(x)=\hat{\tilde{ 
h}}^{(\mathbb{P})}_{\mu \cal N}(x) 
+i\frac{1}{M^2}\partial_{\mu}(\mathbb{P}\hat{\tilde{ 
h}}^{(\mathbb{P})})_{\cal N}(x)\, ,\nn
\eea
satisfying $\partial^{\mu}\hat{\tilde{ 
h}}^{\prime(\mathbb{P})}_{\mu \cal N}(x)=0$. The field combinations 
$(\mathbb{P}\hat{\tilde{ 
h}}^{(\mathbb{P})})_{\cal N}$ play the role of eaten  Goldstone fields to 
provide 
the physical degrees of freedom. 
Let us analyze them in terms of component fields. Using 
(\ref{GenPertMetEffective}), (\ref{ae}) can be decomposed into 
graviton, antisymmetric tensor and vector field polarization tensors as 
\bea
k^{\mu}[{ \tilde
h^{(\mathbb{P})}}_{\mu\nu}(k)-\frac{1}{M^2}k_{\mu}(\mathbb{P}\cdot 
A^{(\mathbb{P})}(k))_{\nu}]&=&0\, ,\nn\\
k^{\mu}[{b^{(\mathbb{P})}}_{\mu\nu}(k)+\frac{1}{M^2}k_{\mu}(\mathbb{P}\cdot 
\overline{\cal M}\cdot 
A^{(\mathbb{P})}(k))_{\nu}]&=&0\, ,\label{hgcond}\\
k^{\mu}[A^{(\mathbb{P})}_{\mu}{}^{N}(k)-\frac{1}{M^2}k_{\mu}(\mathbb{P}\cdot 
\overline{\cal M}\cdot \tilde
h^{(\mathbb{P})}(k)\cdot\overline{\cal M})^{N}]&=&0\, .\nn
\eea

\noindent
{\bf Gravitons}

The first equation in (\ref{hgcond}) can be recast as 
\bea
k^{\mu}\{{ \tilde
h^{(\mathbb{P})}}_{\mu\nu}-\frac{1}{M^2}[k_{\mu}(\mathbb{P}\cdot 
A^{(\mathbb{P})})_{\nu}+k_{\nu}(\mathbb{P}\cdot A)_{
\mu }+k_{\nu}k_{\mu}\frac{1}{M^2}(\mathbb{P}\cdot\overline{\cal M}\cdot \tilde
h^{(\mathbb{P})}.\overline{\cal M}\cdot\mathbb{P})]\}=0\, ,
\nn
\eea
where we have used the third equation in (\ref{hgcond}).
Thus, we have an effective symmetric tensor with polarization $\tilde h'^{(\mathbb 
P)}_{\mu\nu}$ 
satisfying 
\begin{equation}
 k^{\mu}\tilde h^{'(\mathbb{P})}_{\mu\nu}(k)= 0\, ,
\end{equation}
where 
\bea
\tilde  h_{\mu\nu}^{\prime(\mathbb{P})}(k)={ \tilde
h^{(\mathbb{P})}}_{\mu\nu}-\frac{1}{M^2}[k_{\mu}(\mathbb{P}\cdot 
A^{(\mathbb{P})})_{\nu}+k_{\nu}(\mathbb{P}\cdot A^{(\mathbb{P})})_{
\mu }-k_{\nu}k_{\mu}\frac{1}{M^2}(\mathbb{P}\cdot\overline{\cal M}\cdot \tilde
h^{(\mathbb{P})}\cdot \overline{\cal M}\cdot\mathbb{P})]~~~
\label{physgrav}
\eea
is constructed  from the original graviton polarization tensor, one vector 
field   $ (\mathbb{P}\cdot A)_{\nu}$ and a scalar field $\mathbb{P}
\cdot \overline{\cal M}\cdot \tilde h\cdot \overline{\cal M}\cdot\mathbb{P}$, as 
expected from the above counting of degrees of freedom.

\noindent
{\bf Antisymmetric tensor }

We can proceed similarly with the antisymmetric field.
Namely, the second equation in (\ref{hgcond}) can be rewritten as
\bea
 k^{\mu}\{{ 
b^{(\mathbb{P})}}_{\mu\nu}+\frac{1}{M^2}[k_{\mu}(\mathbb{P}\cdot \overline{\cal M}\cdot 
A^{(\mathbb{P})})_{\nu}-k_{\nu}(\mathbb{P}\cdot \overline{\cal M}\cdot 
A^{(\mathbb{P})})_{
\mu }]\}+
\frac{1}{M^2}k_{\nu}k^{\mu}(\mathbb{P}\cdot\overline{\cal M}\cdot A^{(\mathbb{P})})_{
\mu }=0\, ,\nn
\eea
and using the third equation in (\ref{hgcond}),  the last term reads
\bea
k^{\mu}(\mathbb{P}\cdot\overline{\cal M}\cdot A^{(\mathbb{P})})_{
\mu }= - \mathbb{P}\cdot \overline{\cal M}\cdot h^{(\mathbb{P})}\cdot \mathbb{P}\, .
\eea
However, this term  vanishes at first order\footnote{In fact, this can be easily 
seen by rewriting the condition $\mathbb{P}^2=0$. Namely
\small{\begin{eqnarray}
 \mathbb{P}^2&=&\mathbb{P}^{M}{\mathcal{M}}_{M N}
 \eta^{N K}{\mathcal{M}}_{K L}\mathbb{P}^{L}
 =\mathbb{P}^{M}\big(\overline{\mathcal{M}}_{M N} + 
 \tilde h_{M N} \big)\eta^{N K}\big(\overline{\mathcal{M}}_{K L}
 + \tilde h_{K L}\big)\mathbb{P}^{L}\\\nn
&=&\mathbb{P}^2 + 2\mathbb{P}\cdot\overline{\mathcal{M}}\cdot \tilde h\cdot\mathbb{P} + 
\mathcal{O}\big(\tilde h^{2}\big)\nn
\end{eqnarray}}
and, therefore $\mathbb{P}\cdot\overline{\mathcal{M}}\cdot \tilde h\cdot\mathbb{P}=0$}, 
and
then  we are left with an effective antisymmetric polarization
\bea
b'^{(\mathbb{P})}_{\mu\nu}=
b^{(\mathbb{P})}_{\mu\nu}+\frac{1}{M^2}[k_{\mu}(\mathbb{P}\cdot\overline{\cal M}\cdot 
A^{(\mathbb{P})})_{\nu}-k_{\nu}(\mathbb{P}\cdot \overline{\cal M}\cdot 
A^{(\mathbb{P})})_{
\mu }]\, ,
\label{physb}
\eea
where the original polarization $b^{(\mathbb P)}_{\mu\nu}$  ``eats'' a vector 
$ (\mathbb{P}\cdot\overline{\cal 
M}\cdot A^{(\mathbb P)})_{\nu}$, 
in agreement with the discussion above. 

 \noindent
{\bf Vectors}

The third equation (\ref{hgcond}) directly tells us that there are massive 
vector polarizations
\bea
A^{'(\mathbb{P})N}_{\nu}(k)=A_{\nu}^{(\mathbb{P})N}(k)+\frac{1}{M^2} 
k_{\nu}(\mathbb{P}\cdot \overline{\cal M}\cdot \tilde h^{(\mathbb{P})}\cdot 
\overline{\cal M})^{N}+\dots\, ,
\eea
satisfying $k^{\nu}A_{\nu}^{\prime (\mathbb{P})N}=0$.

Thus, from  the $2n$ original vectors $A^{(\mathbb P)N}_{\mu}$, the combination 
$\mathbb{P}\cdot A^{(\mathbb P)}_{\mu}$ is eaten by the graviton and the combination 
$\mathbb{P}\cdot\overline{\cal M}\cdot A^{(\mathbb P)}_{\mu}$ is eaten by the $b^{(\mathbb P)}_{\mu\nu}$ field to become 
massive, and we are 
left with $2n-2$ vectors.
These vectors become massive  by absorbing $2n-2$ scalars 
from the  $n^2$ original $\tilde h^{(\mathbb P)}_{MN}$. One more scalar 
(the combination 
$\mathbb{P}\cdot\overline{\cal M}\cdot \tilde h^{(\mathbb P)}\cdot\overline{\cal M}\cdot\mathbb{P}$) 
is eaten by the graviton, so 
finally we are left with $n^2-(2n-2)-1=(n-1)^2$ scalars.

Notice that the vector eaten by the graviton should be different from the one 
eaten by $b^{(\mathbb P)}_{\mu\nu}$.
Indeed,  this appears to be the case.
If ${\mathbb P}\cdot A^{(\mathbb P)}$ selects some combination, then ${\mathbb P}
\cdot\overline{\cal M}\cdot A^{(\mathbb P)}$ selects an independent one. Actually, 
$\overline{\cal M}$ 
acts effectively by changing lower to upper indices (see (\ref{etaeme})).

The physical states found above should be interpreted from the
generalized gauge transformations. 
Starting with  generic states, there should be a choice of gauge 
parameters $\xi^{\cal M}=({\xi}^{\mu},\tilde {\xi}_{\mu}, \Lambda^M) $ such 
that, by performing a generalized transformation of 
the form (\ref{gendif}), unphysical states are  gauged away.
Let us show that this is indeed the case.
 The generalized diffeomorphisms (\ref{gendifh}), in terms of 
 component fields and up to first order in fluctuations, read 
  \bea
  \delta_{\xi}\tilde  h_{\mu\nu} &= &\partial_{\mu}\xi^{\lambda}\eta_{\lambda\nu} + 
\partial_{\nu}\xi^{\lambda}\eta_{\lambda\mu}\, ,\\
 \delta_{\xi} b_{\mu\nu}&=& 
\partial_{\mu}\tilde{\xi}_{\nu}-\partial_{\nu}\tilde{\xi}_{\mu}\, ,\\
\delta_{\xi} A_{\mu}^N &= &\partial_{\mu}\Lambda^{N} + 
\eta_{\lambda\mu} \overline{\cal M}^{N\cal M}\partial_{\cal M}\xi^{\lambda} - 
\partial^{N}\tilde{\xi}_{\mu}\, ,\\
\delta_{\xi} \tilde h_{MN} &=& \overline{\cal M}_{MP}\partial_{N}\Lambda^{P} +  
\overline{\cal M}_{PN}\partial_{M}\Lambda^{P} -
\overline{\cal M}_{MP}\partial^{P}\Lambda_{N}- 
\overline{\cal M}_{PN}\partial^{P}\Lambda_{M}\, .
\label{physstatesgen}
\eea
In terms of GKK modes, the gauge transformed fields will be
 \bea\nn 
 \tilde h'^{(\mathbb P)}_{\mu\nu}&=&\tilde h^{(\mathbb P)}_{\mu\nu}+\delta_{\xi}\tilde 
 h^{(\mathbb P)}_{\mu\nu} =\tilde h^{(\mathbb P)}_{\mu\nu}
+ik_{(\mu}\eta_{\nu)\lambda}{\xi}^{\lambda(\mathbb P)}\, ,
\nn\\
 b'^{(\mathbb P)}_{\mu\nu}&=&b^{(\mathbb P)}_{\mu\nu}+\delta_{\xi}b^{(\mathbb P)}_{\mu\nu}=
b^{(\mathbb P)}_{\mu\nu}
 +ik_{[\mu}\tilde{\xi}^{\mathbb P)}_{\nu]}\, 
\nn\\
A'^{(\mathbb P)N}_{\mu}&=&A_{\mu}^{(\mathbb P)N}+
\delta A_{\mu}^{(\mathbb P)N} = A_{\mu}^{(\mathbb P)N}+ik_{\mu}\Lambda^{{(\mathbb P)}N} + 
i \eta_{\lambda\mu} (\overline{\cal M}\mathbb{P})^{N}\xi^{{(\mathbb P)}\lambda} - 
i\mathbb{P}^{N}\tilde\xi^{(\mathbb P)}_{\mu}\, ,\nn\\
\tilde  h^{'(\mathbb P)}_{MN}&=& \tilde h^{(\mathbb P)}_{MN}+\delta_{\xi} \tilde h^{(\mathbb P)}_{MN} =
 i (\overline{\cal M}\Lambda^{(\mathbb P)}_{M})\mathbb{P}_N  + i (\overline{\cal M}\Lambda^{(\mathbb P)}_{N})\mathbb{P}_M
-i(\overline{\cal M}\mathbb{P})_{M} \Lambda^{(\mathbb P)}_{N}
-i(\overline{\cal M}\mathbb{P})_{N} \Lambda^{(\mathbb P)}_{M}\, .\nn\\
\label{physfields}
\eea
In order to fix the gauge parameters, we first impose the conditions
\bea
\mathbb{P}\cdot A'^{(\mathbb P)}_{\mu}&=&0\, ,\\
\mathbb{P}\cdot\overline{\cal M}\cdot A^{'(\mathbb P)}_{\mu}&=&0\, ,
\label{unphysA}
\eea
as it should, since the first equation corresponds to the combination eaten 
by the massive graviton and the second one to the combination eaten by
 the antisymmetric tensor $b_{\mu\nu}^{(\mathbb P)}$. These conditions  fix the 
 form of $ \Lambda^{{(\mathbb P)}}_N,{\xi}^{(\mathbb P)\lambda} $
 up to a coefficient and $\tilde{\xi}^{(\mathbb P)}_{\nu}$. By also requiring that 
 \begin{equation}
  \mathbb{P}\cdot\overline{\cal M}\cdot \tilde
h'^{(\mathbb P)}\cdot\overline{\cal M}\cdot\mathbb{P}=0\, ,
\label{unphysh}
 \end{equation}
 since this is the scalar absorbed by the graviton,  this coefficient is fixed. 
 Finally, the gauge transformations required to gauge away the non-physical fields read
 \bea
\xi^{{(\mathbb P)}\lambda} &=& i\frac{1}{M^2}\eta^{\lambda\mu}[(\mathbb{P}
\cdot A^{(\mathbb P)}_{\mu})
-k_{\mu}\frac{1}{ 2{M^2}} (\mathbb{P}\cdot \overline{\cal M}\cdot \tilde h^{(\mathbb P)}
\cdot\overline{\cal M}\cdot\mathbb{P})]\, ,\nn\\
\tilde{\xi}^{(\mathbb P)}_{\mu}&=& -i\frac{1}{M^2} \mathbb{P}\cdot\overline{\cal M}\cdot
A^{(\mathbb P)}_{\mu}\, ,\nn\\
 \Lambda^{(\mathbb P)N} &=& i\frac{1}{  {M^2}} [(\mathbb{P}\cdot\overline{\cal M}
 \cdot h^{(\mathbb P)}\cdot{\cal 
M})^N- \frac{1}{ 2{M^2}}(\overline{\cal M}\mathbb{P})^{N}
\mathbb{P}\cdot\overline{\cal M}\cdot 
h^{(\mathbb P)}\cdot\overline{\cal M}\cdot\mathbb{P}]\, .
\eea

By noticing that  $\overline{\cal M}\cdot \mathbb{P}\cdot\Lambda^{(\mathbb P)N}=0$ and 
$\mathbb{P}\cdot\Lambda^{(\mathbb P)N}=-\frac{1}{ 2{M^2}}\mathbb{P}
\cdot\overline{\cal M}\cdot \tilde
h^{(\mathbb P)}\cdot\overline{\cal M}\cdot\mathbb{P}$, and using the LMC 
(${\mathbb P}^2=0$), it 
is easy to check that (\ref{unphysA}) and 
(\ref{unphysh}) are satisfied.
By replacing these gauge parameters in (\ref{physfields}), we obtain the 
explicit expressions in terms of the old fields.
The resulting physical fields  $\tilde h'^{(\mathbb P)}_{\mu\nu}, b'^{(\mathbb 
P)}_{\mu\nu}$ are the ones given in
 (\ref{physgrav}) and (\ref{physb}), respectively.

Also, 
\begin{eqnarray}\nn
\tilde  h^{'(\mathbb P)}_{MN}&=
&-(\mathbb{P}\cdot \overline{\cal M}\cdot \tilde h^{(\mathbb P)})_M\mathbb{P}_N\nn\\
&&+\frac{1}{ {M^2}}(\overline{\cal M}
\mathbb{P})_{M} [\mathbb{P}\cdot \overline{\cal M}\cdot \tilde
h^{(\mathbb P)}
\cdot\overline{\cal M})_N   - 
\frac{1}{ 2{M^2}}(\overline{\cal M}\mathbb{P})_{N} 
\mathbb{P}\cdot \overline{\cal M}\cdot
\tilde h^{(\mathbb P)}\cdot\overline{\cal M}\cdot\mathbb{P}]
+ M \leftrightarrow N \nn
\end{eqnarray}
 and 
\bea
{A'}^{(\mathbb P)N}_{\mu} &=& A_{\mu}^{(\mathbb P)N}
-k_{\mu}\frac{1}{ {M^2}}
(\mathbb{P}\cdot\overline{\cal M}\cdot \tilde h^{(\mathbb P)}\cdot\overline{\cal M})^N\nn\\
&&- \frac{1}{M^2}(\overline{\cal M}\mathbb{P})^{N}[
 (\mathbb{P}\cdot A^{(\mathbb P)}_{\mu})-\frac{1}{M^2}k_{\mu}(\mathbb{P}\cdot\overline{\cal M}
 \cdot \tilde h^{(\mathbb P)}\cdot \overline{\cal M}\cdot\mathbb{P})]\\\nn
&&- \frac{1}{M^2}\mathbb{P}^{N}
(\mathbb{P}\cdot\overline{\cal M}\cdot A^{(\mathbb P)}_{\mu})\, .
\label{aprima}
\eea

Interestingly enough, since in the harmonic gauge 
 \bea
 k\cdot A^{(\mathbb P)N}= -(\mathbb{P}\cdot\overline{\cal M}\cdot \tilde h^{(\mathbb P)}
 \cdot \overline{\cal M})^N\, ,
 \eea
 then the physical vectors satisfy
 \be
 k\cdot A^{'{(\mathbb P)}N}= 0\, .
 \ee

 Moreover, it can also be  checked that the  physical fields 
 ${A'}^{(\mathbb P)N}_{\mu}, \tilde h'^{(\mathbb P)}_{\mu\nu}, 
 b'^{(\mathbb P)}_{\mu\nu},\tilde h'^{(\mathbb P)}_{MN}$ are invariant under 
generic 
 linearized diffeomorphisms  
 $\delta_{{\xi}^{(\mathbb P)}}=(\delta_{{\xi}^{\mu (\mathbb P)}},
\delta_{{\tilde {\xi}}_{\mu}^ {(\mathbb P)}},\delta_{ \Lambda^{M {(\mathbb 
P)}}})
$ 
 as given in  
 (\ref{physfields}). This is due to the fact that these combinations correspond 
to physical fields.
 The  situation is  analogous in electromagnetism where the electric and 
magnetic fields are 
 a  gauge invariant combination. 
 Here ${A'}^{(\mathbb P)N}_{\mu}, \tilde h'^{(\mathbb P)}_{\mu\nu},  
b'^{(\mathbb P)}_{\mu\nu},
 \tilde h'^{(\mathbb P)}_{MN}$ would be the physical combinations for the 
internal 
 symmetries (symmetries associated to  $\delta_{{\xi}^{(0)}}$ must still be 
fixed).

  Finally let  us discuss the splitting of the symmetric tensor into a 
traceless 
 part and a trace contribution. 
 Of course the splitting can be performed just by adding and subtracting the 
 trace. Let us consider a trace contribution of the form $\tilde 
h_{\mu\nu}^{\phi (\mathbb{P})}=\tilde h _{\lambda}^{\prime 
\lambda(\mathbb{P})}(k)
 \epsilon^{\phi}_{\mu\nu}(k)$ with 
$
  \epsilon^{\phi}_{\mu\nu}(k)=f_d(\mathbb{P})(\eta_{\mu\nu}+k_{\mu}\chi_{\nu}^{ 
(\mathbb{P})}+
 k_{\nu}\chi_{\mu}^{ (\mathbb{P})})
 $,
where we have used the freedom of including a diffeomorphism parameter 
$\chi_{\nu}$
and $f_d $ is a numerical factor (different for massive and massless states). 
The parameters $\chi_{\nu}$ are chosen 
such that $k^{\mu}\epsilon^{\phi}_{\mu\nu}(k)=0$. 
For the massless modes, this leads to  the requirement $k_{\mu}\chi^{ 
(0)\mu}=-1$,
whereas for massive modes with $M^2= \mathbb{P}\cdot {\overline {\cal 
M}}\cdot\mathbb{P}$,
we find $\chi_{\mu}^{ (\mathbb{P})}=\frac{1}{2M^2}k_{\mu}$.
Therefore, the polarization tensor for the traceless symmetric graviton  is 
\bea
 \tilde h_{\mu\nu}^{' G (\mathbb{P})}(k)=\tilde h_{\mu \nu}^{' (\mathbb{P})}(k)
 - \tilde h_{\mu}^{\prime \mu(\mathbb{P})}(k)\epsilon^{\phi}_{\mu\nu}(k)\, ,
\eea
with $f_d=\frac{1}{d-2}$ for the massless modes and $f_d=\frac{1}{d-1}$ for the
massive ones.

However, we still have the freedom to fix the trace $\tilde 
h'^{\lambda}{}_{\lambda}$. 
 A convenient choice is $Tr(\tilde h')=\tilde h'^{\lambda}{}_{\lambda}= 4 \phi$, 
where  $\phi$ is the dilaton 
 field, which amounts to setting $\hat d=0$ (see \ref{cgdilaton}). 
 
 Actually, in order to compare with string theory results, it proves useful to 
redefine the dilaton as
 $\phi^{' (\mathbb{P})}= \sqrt{f_d} \phi^{ (\mathbb{P})}$, and therefore the 
dilaton polarization 
 becomes
  \begin{equation}
  \epsilon^{\phi'}_{\mu\nu}(k)=\sqrt{f_d}(\mathbb{P})
  (\eta_{\mu\nu}+k_{\mu}\chi_{\nu}^{ (\mathbb{P})}+
 k_{\nu}\chi_{\mu}^{ (\mathbb{P})})\, .
 \label{dilationpol}
 \end{equation}
  It is normalized as
  $ \epsilon^{\phi'}(\mathbb{P})\cdot\epsilon^{\phi'}(\mathbb{P})=1$ and also  
$ \epsilon^{\phi'}(\mathbb{P})\cdot \epsilon^G(\mathbb{P})=0$, by construction.
 
 We notice that the  choice $\hat d=0$ eliminates the 
 last  $\hat d$ dependent term from the Lagrangian 
 (\ref{ldft1}). However the dilaton  part is now included 
 in the previous terms due to the splitting 
 $ \tilde h'_{\mu\nu}=\tilde h^{'G}_{\mu\nu}+\tilde h_{\mu\nu}^{'\phi}$. 
 
 Finally, the cubic order Lagrangian to be considered is  
\bea
L_{DFT}= -\frac12 \hat{\tilde  h}'^{{\cal M} {\cal  N}}\partial_{ \cal 
M}\hat {\tilde h}'^{{\cal K}  {\cal L}}\partial_{ 
\cal K}\hat
{\tilde h}'_{\cal NL}+ \frac18  \hat{\tilde h}'^{{\cal M} 
{\cal N}}
 \partial_{\cal M } \hat {\tilde h}'^{{\cal K} {\cal L}} \partial_{\cal 
N} \hat{\tilde {h}}'_{\cal KL} \, ,
\label{ldftphys}
\eea
where only the physical fields identified above must be considered.

 Recall that, even if diffeomorphisms have been used in order to fix the 
physical degrees of freedom, the expression of the action in the harmonic gauge  
can still be 
used since these transformations, up to first order in the fields and on shell, 
do preserve the gauge.
 More explicitly, $\partial_{\mathcal{M}} 
\mathcal{H}^{\mathcal{M} \mathcal{L}}$ changes as  $\partial_{\mathcal{M}} 
\mathcal{H}^{\mathcal{M} \mathcal{L}} \longrightarrow 
\partial_{\mathcal{M}} \mathcal{H}^{\mathcal{M} \mathcal{L}} + 
\delta_{\xi}\big( 
\partial_{\mathcal{M}} \mathcal{H}^{\mathcal{M} \mathcal{L}} \big)$, where from 
(\ref{gendifh})  we read that 
\bea
\delta_{\xi}\big( \partial_{\mathcal{M}} \mathcal{H}^{\mathcal{M} \mathcal{L}} 
\big) &= &\partial_{\mathcal{P}}\xi_{\mathcal{M}} \partial^{\mathcal{P}} 
\mathcal{H}^{\mathcal{M} \mathcal{L}} - 
2\partial_{\mathcal{M}}\partial_{\mathcal{P}}\xi^{\mathcal{M}}\mathcal{H}^{
\mathcal{L} \mathcal{P}} 
-2\partial_{\mathcal{M}}\partial_{\mathcal{P}}\xi^{\mathcal{L}}\mathcal{H}^{
\mathcal{M} \mathcal{P}}\nn\\
&&+~2\partial_{\mathcal{M}}\partial^{\mathcal{M}}\xi_{\mathcal{P}}
\mathcal{H}^{\mathcal{L} 
\mathcal{P}}+2\partial_{\mathcal{M}}\partial^{\mathcal{L}}\xi_{\mathcal{P}}
\mathcal{H}^{\mathcal{M} \mathcal{P}}\, .\nn
\eea
 Since the gauge parameters
$\xi^{\mathcal{M}}$ are already first order in the fields, we obtain (using LMC)
\begin{equation}
\delta_{\xi}\big( \partial_{\mathcal{M}} \mathcal{H}^{\mathcal{M} \mathcal{L}} 
\big) = - 2\bar{\mathcal{H}}^{\mathcal{L} 
\mathcal{P}}\partial_{\mathcal{P}}\partial_{\mathcal{M}}\xi^{\mathcal{M}} 
-2\bar{\mathcal{H}}^{\mathcal{M} 
\mathcal{P}}\partial_{\mathcal{M}}\partial_{\mathcal{P}}\xi^{\mathcal{L}}
+2\partial^{\mathcal{L}}\big(\bar{\mathcal{H}}^{\mathcal{M} 
\mathcal{P}}\partial_{\mathcal{M}}\xi_{\mathcal{P}}\big)\, .
\end{equation}

The second term vanishes due to the e.o.m (see (\ref{eom})) and it can be easily checked that
 $\partial_{\mathcal{M}}\xi^{\mathcal{M}}$ and 
$\bar{\mathcal{H}}^{\mathcal{M} \mathcal{N}} 
\partial_{\mathcal{N}}\xi_{\mathcal{M}}$  are identically 
zero for  the $\xi^{\mathcal{M}}$ parameters found above.  
Thus,  the harmonic gauge does not completely fix the gauge freedom, and we 
can still use the remaining symmetries to gauge away the Goldstone bosons.

\subsection{Unbroken symmetries}
 The fact that physical fields can be defined by absorbing ``Goldstone like 
 fields'' is associated to the spontaneous 
 symmetry breaking by the background.  This issue has been extensively discussed in the 
 literature about KK compactification (see for instance \cite{dolan,KK, Salam:1981xd}). 
 
 Actually, most of the generalized diffeomorphisms are spontaneously broken by the choice of vacuum,
namely
\begin{equation}
\begin{aligned}
\langle g_{\mu \nu}\rangle &= \eta_{\mu\nu},\\\nn
\langle A^{M}_{\mu} \rangle=\langle b_{\mu \nu}\rangle&= 0\, ,\nn\\
\langle {\cal H}_{MN } \rangle &= \overline{\cal M}_{MN}\, .
\end{aligned}
\label{vacuum}
\end{equation}

In fact, only  the zero modes $\xi_{\cal M}^{(0)}(x)$ parametrize the
 local symmetries, whereas the transformations associated to non-zero modes  
$\xi_{\cal M}^{(\mathbb P)}$ 
 are spontaneously broken \footnote{The algebra of diffeomorphisms  is discussed 
in the Appendix.}.
Thus, for instance, the generalized 
internal diffeomorphism parameter $\Lambda^M(x)$ becomes the $U(1)_M$ gauge 
parameter,
under which the physical fields transform as 
 \begin{equation}
 \begin{aligned}
 \delta g^{\mu \nu} &= \Lambda^{M}\partial_{N} g^{\mu \nu}\, ,\\
 \delta A^{N} &= \Lambda^{M}\partial_{M}A^{N} + d\Lambda^{M}\, ,\\
 \delta b &= \Lambda^{M}\partial_{M}b + \frac{1}{2} A^{M}\wedge d\Lambda_{M}\,\ ,\\
 \delta \mathcal{M}_{M N} &= \Lambda^{P}\partial_{P}\mathcal{M}_{M N}\, ,
 \end{aligned}
 \end{equation}
 where these  equations must be understood as holding  for all GKK modes. Recall that 
 $\partial_{M}\Lambda^{N}=0$, i.e.  the gauge parameters
do  not depend on the internal coordinates,  and   $d\Lambda_{M}=0$
for massive modes.
For massless  modes $ A^{(0)N}$, the usual gauge transformations are obtained.
The gauge transformation of the two-form field $b$ is particularly interesting 
since it involves the vector bosons and, for massless fields, it 
gives rise to the familiar Chern Simons three-form.
Actually, there exist two  simple  covariant combinations of fields under the above 
gauge transformations, namely
\begin{equation}
\begin{aligned}
H&=\big( d - A^{M}\wedge\partial_{M}\big) b + \frac{1}{2} A^{M}\wedge \big( d - 
A^{N}\wedge\partial_{N}\big) A_{M}\, ,\\
B_{M}&= \partial_{M}b + \frac{1}{2}A^{N}\wedge\partial_{M}A_{N}\, ,
\end{aligned}
\end{equation}
where $H$ and $B_{M}$  are  spacetime three-form  and two-form, respectively.
We will see that fields in the Lagrangian do group into these combinations.

\subsection{Cubic terms and effective action}
Once the  physical states have been identified, we proceed to consider the 
third order action\footnote{An alternative proposal of this action can be found  in \cite{hs}.}  (\ref{ldftphys}). By splitting the indices of 
the fluctuations  into spacetime and internal components,
the  Lagrangian containing only physical fields, reads

\bea
{\cal L}&=&-\frac1{12}H_{\mu\nu\rho}H^{\mu\nu\rho}+\frac14D_{
\mu}g_{\nu\rho}D_{\sigma}g^{\nu\rho}
g^{\mu\sigma}-\frac12D_{\mu}g_{\nu\rho}D_{\sigma}g^{\mu\nu}
g^{\rho\sigma}\nn\\
&&-\frac14{\cal 
M}_{MN}F_{\mu\nu}^MF^{N\mu\nu}+\frac18g^{\mu\nu}D_\mu{\cal M}_{MN}D^\mu {\cal M}^{MN}
\nn\\
&&+\frac14{\cal 
M}^{MN}\partial_Mg_{\mu\nu}\partial_Ng^{\mu\nu}-\frac12
{\cal M}_{MN}
\partial_PA_\mu^M\partial_QA_\nu^Ng^{\mu\nu}{\cal M}^{PQ}+\frac18 {\cal 
M}^{PQ}\partial_P{\cal M}_{MN}\partial_Q{\cal M}^{MN}\nn\\
&&-\frac14{\cal 
M}^{MN}B_{M\mu\nu}B_{N\rho\sigma}g^{\mu\rho}g^{\nu\sigma}-\frac12{\cal 
M}^{MN}B_{M\mu\nu}F_{N\rho\sigma}g^{\mu\rho}g^{\nu\sigma}\nn\\
&&+\frac12{\cal 
M}^{MN}\partial_MA^{P}_{\mu}D_{\nu}{\cal M}_{NP}g^{\mu\nu}-\frac12{\cal 
M}^{MN}\partial^PA_{M\mu}D_{\nu}{\cal 
M}_{NP}g^{\mu\nu}-\frac12\partial^MA^{N}_{\mu}\partial_NA_{M\nu}g^{\mu\nu}\nn\\
&&-\frac12{\cal 
M}^{MN}\partial_M{\cal M}^{PQ}\partial_P{\cal M}_{NQ}+\frac12{\cal M}^{MN}{\cal 
M}^{PQ}\partial_MA_{P\mu}\partial_QA_{N\nu}g^{\mu\nu}\, ,\label{ea}
\eea
where we have  included  cubic interactions plus some  higher order  terms  
required by spacetime diffeomorphism and gauge invariances.

Here 
\bea
F_{\mu\nu}^M&=&D_{[\mu}A_{\nu]}^M\equiv
\partial_{[\mu}A_{\nu]}^M-A_{[\mu}^N\partial_NA_{\nu]}^M\equiv\partial_{\mu}A_{
\nu}^M-\partial_{\nu}A_{\mu}^M
-A_{\mu}^N\partial_NA_{\nu}^M+A_{\nu}^N\partial_NA_{\mu}^M\nn\\
H_{\mu\nu\rho}&=&D_{[\mu}b_{\nu\rho]}-\frac12A^M_{[\mu}D_\nu A_{\rho]M}\nn\\
&&\equiv D_{\mu}b_{\nu\rho}+D_{\nu}b_{\rho\mu}+D_{\rho}b_{\mu\nu}
-\frac12(A^M_{\mu}D_\nu A_{\rho M}+A^M_{\nu}D_\rho A_{\mu M}+A^M_{\rho}D_\mu 
A_{\nu M})\nn\\
&&~~~~~~~~~~~~~~~~~~~~~~~~~~~~~~~~~~
+\frac12(A^M_{\mu}D_\rho A_{\nu M}+A^M_{\nu}D_\mu A_{\rho M}+A^M_{\rho}D_\nu 
A_{\mu M})\nn\\
B_{M\mu\nu}&=&\partial_{M}b_{\mu\nu} + \frac12 
A^{N}_{[\mu}\partial_{M}A_{N\nu]}\nn\\
&&\equiv \partial_{M}b_{\mu\nu} + \frac12 
A^{N}_{\mu}\partial_{M}A_{N\nu}-\frac12 A^{N}_{\nu}\partial_{M}A_{N\mu}\, ,
\eea
and the derivatives are
\bea
D_{\mu}&=\partial_{\mu} - A^{M}_{\mu}\partial_{M}&\, .
\label{covderivop}
\eea
Recall that $g_{\mu\nu}=\eta_{\mu\nu}+h_{\mu\nu}, {\cal 
M}_{MN}={\overline {\cal 
M}_{MN}}+h_{MN}$, etc.

The Lagrangian (\ref{ea}) has a rather compact expression due to the explicit
$O(n,n)$ invariant setting. 
The fields here depend on both space time and 
internal  coordinates and must still be mode expanded in generalized momenta, 
according to (\ref{cmexp}).
Modes  correspond  to physical fields, in terms of which the
contributions acquire a more familiar shape. Recall that, 
 when acting on the field mode 
$({\mathbb P})$, $-i\partial_{M}\rightarrow {\mathbb P}_{M}$  is just the charge 
operator.

The action  contains both kinetic  and cubic interaction terms 
of massless and massive fields. Covariant derivatives and  Chern-Simons terms 
in the antisymmetric tensor field strength appear as usual.
For instance, the derivative
$D_\mu$ in (\ref{covderivop}) leads, when mode expanded and  acting on a generic 
field  $\Phi^{\mathbb({\mathbb P})}(x)$,  to the covariant derivative 
\bea
D_{\mu}\Phi^{\mathbb({\mathbb P})}(x)&=& (\partial_{\mu} 
-iA^{(0)M}_{\mu}{\mathbb P}_M )\Phi^{(\mathbb P)}(x)\, ,
\eea
where 
${\mathbb P}_M$ is the electric charge with respect to the 
$U(1)_M$ gauge field $A^{(0)M}_{\mu}$. 

We know from (\ref{onnmomenta}) that fields charged under 
$(A^{(0)m}_{\mu},{\tilde A}^{(0)}_{m\mu}) $ carry charge 
$(\frac{n_m}{R^{(m)}},\frac{w^m}{\tilde R_{(m)}})$ and therefore,
in order to have integer charge,  field redefinitions 
\bea
A^{(0)m}_{\mu}&\rightarrow & R^{(m)}A'^{(0)m}_{\mu}\, ,\\
{\tilde A}^{(0)}_{m \mu}& \rightarrow & {\tilde R_{(m)}}{\tilde A}'^{(0)}_{m\mu}\, ,
\eea
must be performed. Therefore, by using the standard definition  $-\frac{1}{4g^2_d}$ 
for  the coefficient of the field strength squared term  in the $d$-dimensional 
Lagrangian, we see 
that the corresponding   gauge  and gravitational coupling constants are 
\bea
{{g}'^{(m)2}_d}&=&\frac{2\kappa_d^2}{ {R^{(m)}}^2},\qquad
{{\tilde g'^{2}_{d(m)}}}=\frac{2\kappa_d^ 2}{\tilde R_{(m)}},\qquad
\kappa_d^ 2= \kappa^2 e^{2\overline d}\, .
\eea

Recall that, since the generalized dilaton is $O(n,n)$ invariant,
$\kappa_d$ is invariant, as expected.

The massless modes in the first line of (\ref{ea}) give rise to the extended Hilbert-Einstein action
(\ref{eh}), now in $d$ dimensions.  The second line contains Abelian field strength kinetic terms 
$-\frac14{\overline{\cal 
M}_{MN}}F_{\mu\nu}^{(0)M}F^{(0)N\mu\nu}=-\frac14 F_{\mu\nu}^{(0)\dot M}F_M^{(0)\mu\nu} $ 
 as well as kinetic 
terms  for the scalars.
The third line has the massive terms for gravitons, vectors and scalars. 
For instance, the term for the vector bosons leads to 
\begin{equation}
 -\frac12
{\overline{\cal M}}_{MN}
A_{\mu}^{({\mathbb P})M}A_\nu^{({-\mathbb P})N}g^{\mu\nu}\mathbb{P}{\overline{\cal M}}\mathbb{P}
= -\frac12
A_{\mu M}^{({\mathbb P})}A_\nu^{({-\mathbb P})\dot M}g^{\mu\nu}{\mathbb{P}}^{\dot M}\mathbb{P}_M\, ,
\end{equation}
with $M^2=\mathbb{P}{\overline{\cal M}}\mathbb{P}={\mathbb{P}}^{\dot M}\mathbb{P}_M$ the mass of the vector, etc.

We present the full expanded expression in the case of circle compactification 
in
(\ref{circleeffaction}) below. 

Let us stress that the action (\ref{ea}) is  an effective 
gauge invariant action. The massless sector contains gravity+Kalb-Ramond field+ 
vector bosons + scalars, 
coupled to the corresponding towers of massive fields associated to KK momenta 
as well 
as windings. Propagators, Feynman rules, etc. which are necessary for field 
amplitudes 
computations can be explicitly obtained.  
It provides a generalization of previous constructions  (see for instance 
\cite{Han:1998sg,GRW})
where  KK compactifications of gravity were considered, 
in diverse phenomenological proposals.

For comparison with string theory amplitudes we will be interested in 
the on shell  action.

\section{String theory amplitudes}
\label{sec:String theory amplitudes}

In this section we consider string theory with  constant 
toroidal backgrounds $G_{pn}$ and $B_{mp}$ for the metric and antisymmetric tensor,
respectively.
We analyze the vertex operators  creating  physical states,
discuss the computation of  their three-point functions 
and contrast with the results  obtained in the previous sections from DFT. 
 We restrict to states with left and right moving oscillator numbers $N=\bar N=1$.
 The vertex operators creating these states are analyzed in two different ways:

On the one hand, we 
show that a combination of different vertex operators (associated to vectors, 
two-tensors or scalars) is needed in order to cancel conformal anomalies. 
These combinations can be identified with the expressions determined by the 
generalized harmonic gauge choice 
on the DFT side  and correspond to a worldsheet manifestation of a 
built-in string   Higgs mechanism. 

 On the other hand, consistency requirements on the full vertex operator, once 
the harmonic gauge was chosen, fix the physical polarizations and it is with 
these operators, corresponding to physical degrees of freedom,  that all 
scattering amplitudes are computed.

\subsection{Conformal anomalies and DFT harmonic condition}
 
It is known that the cancellation of conformal anomalies at the string  world 
sheet level manifests as gauge symmetry requirements on the target space fields.
This is indeed the case here. The different vertex operators 
corresponding to two-tensor, vector and scalar fields will generically have 
anomalous OPEs (Operator Product Expansions)  with the world sheet stress energy tensor.
For massless fields, the cancellation of anomalous terms leads to the familiar 
gauge 
conditions $k^{\mu}\epsilon^G_{\mu \nu}(k)= 0,\, 
k^{\mu}\epsilon^M_{\mu}(k)=0$, 
etc. 
for the polarization tensors of  gravitons, vectors, etc. These correspond to 
equations (\ref{hgcond}) for zero 
generalized momentum.

For massive fields,  a combination of the different vertex 
operators must be considered, such that the sum of the different anomalous 
contributions  cancel. This is, indeed, a world sheet manifestation of 
the Higgs mechanism. The conditions for cancellation of the anomalous terms can 
be written in an $O(n,n)$ language and can be shown to coincide with the harmonic 
gauge conditions found  in DFT. 


The vertex operators we are interested in are, up to normalizations,  
\begin{eqnarray}\nn
 V_{G}&=& \epsilon^{G}_{\rho\sigma }(k, k_L, k_R):\partial X^\rho \bar\partial  
  X^\sigma \, e^{ik\cdot X+i
    k_L\cdot Y+ik_R\cdot\bar Y}:\, ,\\\nn
  V_{A_R}&=& \epsilon^a_{R\rho}(k, k_L, k_R):\partial X^\rho \bar\partial  
\bar Y ^a\,e^{ik\cdot X+ik_L\cdot Y+ik_R\cdot\bar Y}:\, ,\\\nn
V_{A_L}&=& \epsilon^a_{L\rho}(k, k_L, k_R):\partial 
Y^a\, \bar\partial X^\rho \,e^{ik\cdot X+ik_L\cdot Y+ik_R\cdot\bar Y}:\, ,\\
V_{\phi}&=& \phi_{ab}(k, k_L, k_R):\partial 
Y^a \, \bar\partial \bar Y^b \,e^{ik\cdot X+ik_L\cdot Y+ik_R\cdot\bar Y}:
\label{vop}
\end{eqnarray}

The label $G$ generically denotes a symmetric traceless, antisymmetric or trace polarization, 
$A_L, A_R$ refer to vectors and $\phi$ to  scalars.
 Here $\bar\partial=\partial_{\bar z}, \partial=\partial_z$
and $Y=Y(z), \bar Y=\bar Y(\bar z)$ denote left and right moving coordinates.
It is convenient to use coordinates $Y^a=e_m{}^a Y^m$ with
tangent space indices $a, b, ...$, defined in terms of the vielbein
$e_m{}^a$ ($\delta^{ab}=e_m{}^a g^{mn}e_n{}^b$) since they
have the standard OPEs.
Namely, the propagators read
\bea
\langle X^\mu(z,\bar z) X^\nu(w,\bar w)\rangle &=& -
\frac{\alpha '}2\eta^{\mu\nu}ln|z-w|^2\, ,\nn \\
\langle Y^a (z) Y^b(w)\rangle= -\delta^{ab}
\frac{\alpha '}2 ln(z-w)\, ,&&\qquad
\langle \bar Y^a(\bar z) \bar Y^b(\bar w)\rangle = -\delta^{ab}
\frac{\alpha '}2 ln(\bar z-\bar w)\, .\nn
\eea

The vertex operator momenta
are
\bea
k_{aL}=e_a{}^mp_{mL}\, ,\qquad
k_{aR}=e_a{}^mp_{mR}\, ,
\eea
where
\be
p^m_L=\tilde p^m
+g^{mn}(p_n-B_{nk}\tilde p^k)\, ,\qquad
p^m_R= -\tilde p^m
+g^{mn}(p_n-B_{nk}\tilde p^k)\, .\nn
\ee
 The  stress energy tensor is
 	\bea
 T(z)  &=&-\frac 1{\alpha '}(\eta_{\mu\nu}:\partial 
X^\mu(z)
 \partial X^\nu(z):+\delta_{ab}:\partial Y^a(z)\partial Y^b(z):) 
\, ,\nn
 \eea
and similarly for the right moving one. The OPEs are
\bea \nn
  T(z_1) V_G(z_2)&=& [\frac{\ap}{4}(k^2+k_L^2)+1]\frac{ 
V_G}{z_{12}^2}-2i\frac{\ap}{4z_{12}^3}[:k^\rho\epsilon^G_{\rho\sigma }\partial 
\bar X^\sigma \, e^{ik\cdot X+ik_L\cdot Y+ik_R\cdot\bar Y}:]+\dots\, ,\nn\\
T(z_1)V_{A_L}&=& [\frac{\ap}{4}(k^2+k_L^2)+1]\frac{ 
V_{A_L}}{z_{12}^2}-2i\frac{\ap}{4z_{12}^3}[:k^a_L 
\epsilon^a_{L\rho}\partial \bar X^\rho \, e^{ik\cdot X+ik_L\cdot 
Y+ik_R\cdot\bar 
Y}:]+\dots\, ,\nn\\
 T(z_1) V_{A_R}&=&  [\frac{\ap}{4}(k^2+k_L^2)+1]\frac{ 
V_{A_R}}{z_{12}^2}
-2i\frac{\ap}{4z_{12}^3}[:k^\rho
\epsilon^a_{R\rho}\partial \bar Y^a \, e^{ik\cdot X+k_LY+k_R\bar 
Y}:]+\dots\, ,\nn\\
T(z_1)V_{\phi}&=& [\frac{\ap}{4}(k^2+k_L^2)+1]\frac{ 
V_{\phi}}{ z_{12}^2}
-2i\frac{\ap}{4z_{12}^3}[:k^a_L\phi_{ab}\partial  \bar Y^b \, e^{ik\cdot 
X+ik_L\cdot
    Y+ik_R\cdot\bar Y}:]+\dots\, .\nn
 \eea
  Since $k_L^2=-k^2$,  the 
vertex 
operators have the correct conformal weight $h=1$ (and similarly $\bar h=1$), however, 
there are cubic anomalies which suggest that the physical fields should be 
created from combinations of these operators.
Consider then the vertex associated with the massive graviton
\begin{equation}
 V=\alpha V_G+ \beta V_{A_L}+\gamma V_{A_R}+\delta V_{\phi}\, ,
 \label{vvertexgr}
\end{equation}
with constant $\alpha, \beta, \gamma, \delta$. From the OPE with $T$ and $\bar 
T$, the anomaly cancellation conditions are
\bea
\label{cancon2}
  \alpha k^\rho\epsilon_{\rho\sigma}+\beta k^a_L \epsilon^a_{L\sigma}&=&0\, 
,\qquad
 \alpha k^\rho\epsilon_{\rho\sigma }+\gamma k^a_R   
\epsilon^a_{R\sigma}=0\, ,\nn\\
\delta k^a_L\phi_{ab}+\gamma \epsilon^b_{R\sigma}k^\sigma&=&0\, ,\qquad
 \delta k^a_R\phi_{ba}+\gamma 
k^\rho\epsilon^b_{L\rho}=0\, 
\eea

Choosing $2\alpha=\gamma=\beta$, the sum of the first two equations leads to 
\bea
  k^\rho\epsilon^G_{\rho\sigma}+  k_L^a\epsilon_{La}+ k_R^a\epsilon_{Ra}=
  k^\rho\epsilon^G_{\rho\sigma}+
    \tilde p^m
\tilde  \epsilon_{m\sigma}+g^{mn}p_n
\epsilon_{m\sigma}=k^\rho \tilde h^{( 
\mathbb{P})}_{\rho\sigma}+\mathbb{P}\cdot{A_{\sigma}^{( 
\mathbb{P})}}=0 \, ,
\label{cananomaly1}
\eea
where we have defined
\bea
\epsilon_{m\sigma}&=&\epsilon_{Lm\sigma}+\epsilon_{Rm\sigma},\qquad 
\tilde\epsilon_{m\sigma}+B_{mn}\epsilon^n_\sigma =\epsilon_{Lm\sigma}-\epsilon_
{Rm\sigma}\, ,\label{poldftstrings}
\eea
and we have made the identifications
\bea
{A_{M\sigma}^{( 
    \mathbb{P})}}&=&( A^{(\mathbb P)}_{m\sigma}, 
A^{(\mathbb P)m}_\sigma)\equiv (\tilde\epsilon_{m\sigma},\epsilon^m_{\sigma}),\qquad 
\epsilon^G_{\rho\sigma}\equiv \tilde h^{( 
\mathbb{P})}_{\rho\sigma}
.
\label{identif1}
\eea
Therefore, (\ref{cananomaly1}) is nothing but the first harmonic gauge condition 
in (\ref{hgfields}) in momentum space.

On the other hand, by subtracting the first two equations in (\ref{cancon2}), we 
obtain
\bea
k_L\epsilon_L-k_R\epsilon_R&=&\tilde 
p^m\epsilon_{m\sigma}+g^{mn}(p_n+B_{nk}\tilde p^k)
(\tilde\epsilon_{m\sigma}+B_{mp}\epsilon^p_\sigma)=0\, ,\nn
\eea
which can be written as 
\begin{equation}
 \mathbb{P}\cdot{\overline {\cal M} \cdot A_{\sigma}}^{(\mathbb{P})}=0\, ,
\end{equation}
as found in (\ref{unphysA}).

The other two equations involving the scalars lead to
\bea
\delta (k_L^m
\phi_{mn}+k^n_R\phi_{mn})+\gamma k.(\epsilon^m_L+\epsilon^m_R)&=&
2\delta g^{mn}(p_n+B_{nk}\tilde p^k) \phi_{mn}+
\gamma  k\cdot\epsilon_n=0\, ,
\nn\\
\delta (k_L^m\phi_{mn}-k_R^n\phi_{mn})-\gamma k.(\epsilon_{mL}-\epsilon_{mR})&=&
2\delta \tilde p^m \phi_{mn}-\gamma  
k\cdot(\tilde\epsilon_n+B_{nl}\cdot\epsilon^l)
=0\, ,\nn
\eea
which can be shown to coincide with the third equation of the 
harmonic gauge conditions in (\ref{hgfields}) when choosing $\delta=\frac12 \gamma$
and  establishing the identification with DFT scalar fields (\ref{scalarsmatrix})
\bea\nn
\phi^{m n} + \phi^{n m} &=& \tilde h^{m n}\, ,\\
\phi^{m n} - \phi^{n m} &= & b^{m n} .
\label{polarizationscalarsid}
\eea

Thus, the physical vertex operator for the massive graviton 
is 
\begin{equation}
 V=\frac12\ V_G+  V_{A_L}+ V_{A_R}+\frac 12V_{\phi}.
 \label{vertexgraviton}
\end{equation}
The effective symmetric polarization tensor can be shown to coincide with 
(\ref{physgrav}).

Similar steps can be followed for the Kalb-Ramond field and the second equation in
(\ref{hgfields}) is obtained.

In the next section we introduce
 the physical vertex operators used in the computation of  scattering 
amplitudes. The  anomaly free conditions on the polarizations 
coincide with those of the  physical fields  redefined through the use of 
the harmonic gauge condition.
\subsection{Physical Vertex Operators on the Torus}
\label{subsec:Physical Vertex Operators on the Torus}

In the same way that we found the anomaly free combinations of vertex operators
(or equivalently, the harmonic gauge conditions), we can impose that  each one of the 
vertex operators (\ref{vop})  be  anomaly free. 
This would give the conditions to be satisfied by the physical polarizations, that now 
we distinguish with a prime.  
 Note that this procedure  will give  identically zero
 polarizations for  massive vectors and scalars in the case of only one compact dimension, thus
confirming that there are no such degrees of freedom on a circle compactification.
 
The anomaly cancellation conditions for vectors are
\begin{equation}
\begin{aligned}
k^{a}_{L} \epsilon_{L \rho}'^{a} &= 0 \, ,\qquad
k^{a}_{R} \epsilon_{R \rho}'^{a} &= 0\, , \\
k^{\rho} \epsilon_{L \rho}'^{a} &= 0 \, ,\qquad
k^{\rho} \epsilon_{R \rho}'^{a} &= 0\, .
\end{aligned}
\end{equation}
The first two equations can be combined as
\begin{equation}
\begin{aligned}
k^{a}_{L} \epsilon^{'a}_{L \rho} + k^{a}_{R} \epsilon_{R \rho}^{'a} &= 0 \quad {\rm or~as}\quad
k^{a}_{L} \epsilon_{L \rho}^{'a} - k^{a}_{R} \epsilon_{R \rho}^{'a} &= 0\, ,
\end{aligned}
\end{equation}
which are equivalent to
\begin{equation}
\begin{aligned}
\mathbb P \cdot A^{'}_{\mu} &= 0\,  , \\
\mathbb P \cdot \mathcal{M} \cdot A^{'}_{\mu} &= 0 \, , \\
\partial^{\mu} A^{'B}_{\mu} &= 0\, .
\end{aligned}
\end{equation}
Namely, the conditions found in (\ref{unphysA}) 
after gauge fixing.  
In the same way, for  scalars we find 
\begin{equation}
 k^{a}_{L} \phi^{' a b} = 0\, , \qquad k^{b}_{R} \phi^{'a b} = 0\, ,
\end{equation}
which can be expressed in terms of $\tilde h^{'}_{m n}$ and $b^{'}_{m n}$ as 
the following two 
conditions
\begin{equation}
\begin{aligned}
-\tilde{p}^{m} \tilde h^{'}_{m n} + \tilde{p}^{m} B_{m k} G^{k s} b^{'}_{s n} 
+ p_{m} G^{m k} 
b^{'}_{k n} &= 0\, , \\
-\tilde{p}^{m} b^{'}_{m n} + \tilde{p}^{m} B_{m k}  G^{k s} \tilde h^{'}_{s n} 
+ p_{m} G^{m k} 
\tilde h^{'}_{k n} &= 0\, .
\end{aligned}
\end{equation}
 These coincide with the DFT condition (see \ref{unphysh})
\begin{equation}
\mathbb P \cdot M \cdot \tilde h^{'} \cdot M = 0\, ,
\end{equation}
which represents the Goldstone boson absorbed by the massive vectors.
 
 For the tensors $ \tilde h^{'}_{\mu \nu}$ and $ b^{'}_{\mu \nu}$ we get the usual 
transverse gauge conditions
\begin{equation}
\begin{aligned}
k^{\mu} \tilde h^{'}_{\mu \nu} = 0\, ,\\
k^{\mu} b^{'}_{\mu \nu} = 0\, .
\end{aligned}
\end{equation}
Finally, the dilaton vertex can be written as 
\begin{equation}
V_{\phi}= \phi \epsilon^{\phi}_{\mu \nu}\partial X^{\mu} \bar{\partial} X^{\nu} 
e^{i k\cdot X}\, ,
\end{equation}
with 
\begin{equation}
 \epsilon^{\phi}_{\mu \nu}=\sqrt{f_d}\big( \eta_{\mu \nu} + k_{\mu}\bar{k}_{\nu} + 
k_{\nu}\bar{k}_{\mu} \big)\, ,
\end{equation}
as found in (\ref{dilationpol}) by identifying $\bar{k}_{\nu}\equiv \chi_{\nu}^{(0)}$ 
for the  massless case and $\bar{k}_{\nu}\equiv \chi_{\nu}^{ (\mathbb{P})}$ for massive dilatons.

Thus, we have obtained the requirements that physical polarizations must satisfy. 
 
 Notice that the two approaches to deal with vertex operators provide different  
 information on the theory: The first one displays a built in Higgs mechanism 
 exhibiting the  Goldstone bosons. The  second one deals with the physical degrees 
 of freedom once the gauge was chosen. 
 Of course,  one can obtain the latter using the former, as was shown in the 
 previous section.
We will use physical polarization tensors to compute   scattering amplitudes.

\subsection{Three-point interaction terms} 

In this section we consider  three point functions of the massless and massive 
string states created by the vertex operators described above.
The resulting amplitudes are then compared with the DFT action (\ref{ldftphys}), 
evaluated on shell. We sketch the computation here and provide some details for
the circle case in 
the Appendix.

For the sake of clarity we first concentrate on the  circle compactification.
This case is particularly simple since  neither  physical massive vectors 
nor massive scalars  are present. 
The string S-matrix  three-point amplitudes are presented in 
(\ref{ap:String computations}).  
When mode expanding (\ref{ldftphys}) and  by using the identifications (\ref{identif1}) 
and  (\ref{polarizationscalarsid})     
 between string polarization  tensors and DFT fields polarizations, 
 complete agreement is achieved if we further identify
 \begin{equation}
 \pi g_{c}=\frac{1}{2\kappa^2_d}\, ,
 \end{equation}
 where $g_c$ is the closed string coupling.
 
 The effective $U(1)\times U(1)$ gauge invariant action,
 containing massless as well as 
massive states with these S-matrix elements, can be written down. 
By including terms required from gauge invariance and diffeomorphism invariance,
this action reads
\begin{equation}
 S=\frac{1}{2\kappa^2_d}\int d^{D-1}x \sqrt{-g}{\cal L}
\end{equation}
with
\begin{eqnarray}\nn
{\cal L}&=& R - \frac{1}{12}H_{\mu\nu\rho}^{2} 
-\frac{1}{4}\partial_{\mu}\Phi \partial^{\mu}\Phi  \\\nn
 &-& \frac{1}{4}F_{\mu\nu}F^{\mu\nu} 
-\frac{1}{4}\tilde{F}_{\mu\nu}\tilde{F}^{\mu\nu} 
+\frac{1}{2}F_{\mu\nu}F^{\mu\nu}\Phi -
\frac{1}{2}\tilde{F}_{\mu\nu}\tilde{F}^{\mu\nu}\Phi  \\\nn
 &-&\frac{1}{2}\sum^{\infty}_{n=1}\left(\mathcal{D}_{\rho}
h^{*\,(n)}_{\mu\nu}\mathcal{D}^{\rho}h^{(n)\,\mu\nu} -
2\mathcal{D}_{\mu}h^{*\,(n)}_{\nu\rho}\mathcal{D}^{\nu}h^{(n)\,\mu\rho}
 + m^{2}_{n}h^{*\,(n)}_{\mu\nu}h^{(n)\,\mu\nu} \right)  
\\\nn
 &-&\frac{1}{2}\sum^{\infty}_{w=1}\left(\mathcal{D}_
{\rho}\tilde{h}^{*\,(w)}_{\mu\nu}\mathcal{D}^{\rho}\tilde{h}^{(w)\,\mu\nu
} -
2\mathcal{D}_{\mu}\tilde{h}^{*\,(w)}_{\nu\rho}\mathcal{D}^{\nu}\tilde{
h}^{(w)\,\mu\rho} +
m^{2}_{w}\tilde{h}^{*\,(w)}_{\mu\nu}\tilde{h}^{(w)\,
\mu\nu} \right)  \\\nn
 &+&\sum^{\infty}_{n=1} \left(
\frac{1}{6}|H_{\mu\nu\rho}^{(n)}|^{2} +\frac{m^{2}_{n}}{2}|b^{(n)}_{\mu\nu}|^{2} \right)
+\sum^{\infty}_{w=1}\left(\frac{1}{6}|\tilde{H}_{\mu\nu\rho}^{(w)}|^{2} 
+\frac{m^{2}_{w}}{2}|\tilde{b}^{(w)}_{\mu\nu}|^{2} \right) \\\nn
  &+&\sum^{\infty}_{n=1}\frac{1}{2}\frac{n^{2}}{R^{2}}\left(|h^{
(n)}_{\mu\nu}|^{2} + |b^{(n)}_{\mu\nu}|^{2}\right)\Phi - 
\sum^{\infty}_{w=1}\frac{1}{2}\frac{w^{2}}{\tilde{R}^{2}}
\left(|\tilde
{h}^{(w)}_{\mu\nu}|^{2}+ |\tilde{b}^{(w)}_{\mu\nu}
|^{2}\right)\Phi \\\nn
 &-&i \sum^{\infty}_{n=1}\frac{n}{R}\left(h^{*(n)}_{\mu\nu}b^{(n)\nu}_{
\rho} 
+h^{(n)}_{\mu\nu}b^{*(n)\nu}_{\rho}\right)\tilde{F}^{\mu\rho}
 -i\sum^{\infty}_{w=1}\frac{w}{\tilde{R}}\left(\tilde{
h}^{*(w)}_{\mu\nu}\tilde{b}^{(w)\nu}_{\rho} 
+\tilde{h}^{(w)}_{\mu\nu}\tilde{b}^{
*(w)\nu}_{\rho}\right)F^{\mu\rho}\\\nn
 &+&\sum_{n_{1}+n_{2}+n_{3}=0}^{n_{i}\neq 0}\left(\frac{1}{4} \mathcal{D}_{\mu} 
h^{(n_{1})}_{\rho\sigma}\mathcal{D}_{\nu} 
h^{(n_{2})\,\rho\sigma}h^{(n_{3})\,\mu\nu} 
-\frac{1}{2}h^{(n_{1})}_{\mu\rho}\mathcal{D}^{\mu}h^{(n_{2})}_{\nu\sigma}
\mathcal{D}^{\nu}h^{(n_{3})\,\rho\sigma}\right)\\\nn
  &+&\sum_{w_{1}+w_{2}+w_{3}=0}^{w_{i}\neq 
0}\left(\frac{1}{4} \mathcal{D}_{\mu} 
\tilde{h}^{(w_{1})}_{\rho\sigma}\mathcal{D}_{\nu} 
\tilde{h}^{(w_{2})\,\rho\sigma}\tilde{h}^{(w_{3}
)\,\mu\nu} 
-\frac{1}{2}\tilde{h}^{(w_{1})}_{\mu\rho}\mathcal{D}^{\mu}\tilde{h
}^{(w_{2})}_{\nu\sigma}\mathcal{D}^{\nu} 
\tilde{h}^{(w_{3})\,\rho\sigma}\right)\\\nn
 &+&\sum_{n_{1}+n_{2}+n_{3}=0}^{n_{3}\neq 0} \left(\frac{1}{4}\mathcal{D}_{\mu} 
b^{(n_{1})}_{\rho\sigma}\mathcal{D}_{\nu} 
b^{(n_{2})\,\rho\sigma}h^{(n_{3})\,\mu\nu} 
-\mathcal{D}_{\mu}b^{(n_{1})\,\sigma\nu}\mathcal{D}_{\nu}b^{(n_{2})}_{
\sigma\rho}h^{(n_{3})\,\mu\rho}\right.\\\nn
&&\,\,\,\,\,\,\,\,\,\,\,\,\,\,\,\,\,\,\,\,\,\,\,~~~~~~~-  \left.\frac{1}{2} 
b^{(n_{1})\,\rho\mu}\mathcal{D}_{\mu}b^{(n_{2})\,\sigma\nu}\mathcal{D}_{\nu}
h^{(n_{3})}_{\rho\sigma}\right)\\\nn
 &+&\sum_{w_{1}+w_{2}+w_{3}=0}^{w_{3}\neq 0} \left(
\frac{1}{4}\mathcal{D}_{\mu} 
\tilde{b}^{(w_{1})}_{\rho\sigma}\mathcal{D}_{\nu} 
\tilde{b}^{(w_{2})\,\rho\sigma}\tilde{h}^{(w_{3})\,\mu\nu} 
-\mathcal{D}_{\mu}\tilde{b}^{(w_{1})\,\sigma\nu}\mathcal{D}_{\nu}\tilde
{b}^{(w_{2})}_{\sigma\rho} \tilde{h}^{(w_{3})\,\mu\rho}\right.\\
&&\,\,\,\,\,\,\,\,\,\,\,\,\,\,\,\,\,\,\,\,\,\,\, ~~~~~~- \left.
\frac{1}{2} 
\tilde{b}^{(w_{1})\,\rho\mu}\mathcal{D}_{\mu}\tilde{b}^{(w_{2})\,\sigma\nu}
\mathcal{D}_{\nu} \tilde{h}^{(w_{3})}_{\rho\sigma}\right)
\label{circleeffaction}
\end{eqnarray}
 where $\Phi$ denotes the massless scalar; 
$h^{(n)}_{\mu\nu}$ and $\tilde h^{(
w)}_{\mu\nu}$ the modes of the massive 
graviton with momentum $n$ and winding $w$ respectively; 
$b^{(n)}_{\mu\nu}$ and $\tilde b^{(
w)}_{\mu\nu}$ 
the modes of the massive antisymmetric tensor with momentum $n$ and winding $w$ respectively.

We have introduced  the following definitions
\begin{equation}
\begin{aligned}
&\nabla_{\mu}f_{\rho\sigma}=\partial_{\mu}f_{\rho\sigma} - 
\Gamma^{\lambda}_{\rho\mu}f_{\lambda\sigma} -\Gamma^{\lambda}_{\sigma\mu}f_{\rho d}\, ,\\
&\Gamma^{\lambda}_{\mu\nu}=\frac{1}{2} g^{\lambda\sigma}\left( \partial_{\nu}g_{\sigma\mu} 
+ 
\partial_{\mu}g_{\sigma\nu}-\partial_{\sigma}g_{\mu\nu} \right)\, ,\\
&F_{\mu\nu}=\nabla_{\mu}A_{\nu} - \nabla_{\nu}A_{\mu}\, ,\\
&\mathcal{D}_{\mu}=\nabla_{\mu} -iA_{\mu} \hat{q}_{n}
-i\tilde{A}_{\mu}\hat{q}_{w}\, ,\\
&H_{\mu\nu\rho}= 
\mathcal{D}_{\mu}b_{\nu\rho}+\mathcal{D}_{\rho}b_{\mu\nu}+\mathcal{D}_{\nu}b_{
\rho\mu}\, .
\end{aligned}
\end{equation} 
  
 Here indices are  raised with the inverse of the metric tensor 
$g_{\mu\nu}$,  $\hat{q}$ is the charge operator, complex conjugation is 
denoted  with $*$ and, under charge conjugation, the momentum or winding change 
sign i.e 
$h^{(n)\,*}=h^{(-n)}$.

The kinetic terms of the symmetric massive states produce  the known Fierz 
Pauli Lagrangian \cite{fp,ortin}, and the T-duality symmetry $R\leftrightarrow \tilde R, 
n\leftrightarrow w$ is manifest.
This action coincides with (\ref{ea}) when specified for the 
circle case.

\subsection{Strings vs DFT on generic tori}
 
 Generalizing the results obtained for the circle to generic 
tori is formally straightforward.   However, the number of terms involved is 
much 
bigger.   The massless sector contains, besides the graviton, dilaton, 
antisymmetric and scalar fields,
the $2n$  gauge fields associated to $U(1)^n\times U(1)^n$.   The massive sector
includes now, generically,  massive vectors and scalars.   
  The comparison of DFT cubic interactions contained in the mode expansion of 
  (\ref{ldftphys}) with  three point  scattering amplitudes computed using the 
vertex 
  operators (\ref{vop}) is now performed with the help of the symbolic algebra 
  computer program   XCadabra \cite{{Peeters:2006kp}}. 
  Our algorithm compares three point scattering amplitudes of string states and 
DFT 
 cubic interaction terms  by   systematic use of momentum conservation and on 
shell conditions\footnote{The program is available upon request to the 
authors.}.

 As an example of the calculated quantities, we present the result of the 
scattering amplitude between one antisymmetric tensor $b_{\mu \nu}$ (with 
momentum $k_{1 \, \mu}$, and charges $p_{1 \, m}$ and $w^{1 \, m}$), one vector
$A^{m}_{L \mu}$ (with momentum $k_{2 \, \mu}$, and charges $p_{2 \, m}$ and 
$w^{2 \, m}$) and one antisymmetric scalar $b_{m n}$  (with momentum $k_{3 \, 
\mu}$, and charges $p_{3 \, m}$ and $w^{3 \, m}$). 
 
 In the DFT action there is only one place where the interaction vertex can be 
found, namely
\begin{equation}
- \frac{1}{2\kappa_d^2} \partial_{M} b_{\mu \nu} \partial_{\rho} A_{N 
\sigma}{M}^{MN} 
g^{\mu \rho} g^{\nu \sigma}\, .
\end{equation}  
 Splitting the double internal indices, in order to exhibit the  explicit 
contributions of   $b_{m n}$ 
and $h_{m n}$ scalars, one can collect the  required interactions  and compute 
the three point amplitude. The result is
\begin{eqnarray}\nn
\frac{1}{2\kappa_d^2}
 \epsilon_{\mu \nu}(k_1) \epsilon_{L{\mu m}}(k_2)G^{n m} b_{n k}(k_3)  
\big[ k_{2 \nu} w^{1 k} - 
G^{k s} (k_3)   k_{2 \nu} p_{1 s}
+ B_{sl}  G^{s k} (k_3)  k_{2 \nu}  w^{1 l}\big]\, ,
\end{eqnarray}
where $\epsilon_{\mu \nu}$, $\epsilon_{L \mu m}$ and $b_{n k}$ are the 
polarizations of 
the two-form, the left vector and the scalar, respectively.
 The same result is obtained in string theory if we choose 
$\frac{1}{2\kappa_d^2} =
 \pi g_{c}$.

\section{Conclusions and Outlook}
\label{sec:Conclusions and Outlook}

 Double Field Theory was originally motivated by toroidal compactifications
 and a double set of coordinates was proposed as conjugate variables of compact 
 momenta and windings. However, a specific realization of momentum and winding 
modes,
which generically requires dealing with massive states, was lacking.

In this work we have dealt with  massless and massive states   of  DFT  
compactifications on generic  double tori  (in presence of constant background 
fields) and compared them  with  a slice of the massless and 
massive  states of bosonic 
string  theory  compactified on a torus.
The slice considered corresponds to states with excitation numbers $N=\bar 
N=1$, 
namely, a subsector of the 
bosonic string arising from states containing one left and one right moving 
oscillators. 

We found complete agreement between the spectra of both DFT and string theory 
when a level matching 
constraint is imposed on the DFT side. Moreover, by expanding the generalized 
fields of DFT at first order    in  fluctuations around the 
constant background,  the resulting third order action  agrees with the 
effective 
action arising from  three-point scattering amplitudes in string theory.
For $n$ dimensional tori and  $d$ space-time  dimensions the obtained action 
corresponds to a 
gauge theory with $G_n=U(1)^n\times U(1)^n$ Abelian gauge group coupled to 
gravity. 
The computations involve both KK and winding modes, named here GKK modes, and 
therefore the 
action contains an infinite number of charged  massive fields. 

It is worth emphasizing that  DFT provides a concise and manifestly $O(n,n)$ 
realization of this effective string theory action.   Moreover, on a 
$2n$-dimensional double torus background, the global $O(n,n,\mathbb R)$ 
symmetry 
of DFT is broken to $O(n,n, {\mathbb Z})$, the discrete 
 T-duality group of the full string theory.

As is well known, physical states in string theory are selected by ensuring  
cancellation of
conformal anomalies in the world sheet. We found that  the DFT 
manifestation of these requirements is the invariance under generalized 
diffeomorphisms. By 
using such  invariance, we have shown that a generalized harmonic gauge 
condition 
can be chosen, and established a correspondence with conditions derived from 
string 
theory. Interestingly enough, this gauge choice allows to identify the 
different 
Goldstone modes that are absorbed to generate physical fields.
Besides the gravity multiplet and massless vectors associated to the 
compactified gravitational and 
antisymmetric fields, physical massive fields correspond to massive 
symmetric and antisymmetric tensors, vectors and scalars charged under the $ 
G_n$ gauge group. The charges, corresponding to momentum and winding 
numbers, are simply encoded in the generalized DFT  momenta $\mathbb{P}$.
Generalizing  known results  in KK compactifications, we found 
the  infinite global symmetry algebra associated to infinite local generalized 
parameters. In particular, it contains a finite 
Poincar\'e   $  \times SO(1,2)^n \times  SO(1,2)^n$  subalgebra and  
massive states should organize in its (infinite dimensional) representations.

Of course the effective action reproducing the three-point amplitudes of these 
physical massless
 and massive string states 
  is not a low energy effective action since all 
possible massive levels are involved. The action provides an organized 
truncation of  string theory. However this truncation is incomplete since it 
contains states with masses of the order or higher than those of string 
states with $N$ and/or $\bar N\ne 1$ that were not included here.
Indeed, we  know from string theory that  
new fields involving higher spins (associated with $N$ and/or $\bar N\ne 1$)
 appear in the spectrum and play a crucial role in higher point-amplitudes. In 
DFT language, higher order $O(n,n)$ 
generalized tensors, incorporating these missing string degrees of freedom, are 
expected.

We also know that a gauge symmetry enhancing,  associated 
to the presence of windings, occurs in string theory at self dual points. This 
enhancing involves 
states with $N-\bar N\ne 0$  (e.g. $N, \bar N=0,\pm 1$) and for this reason  it 
cannot be seen in our construction.  In \cite{aimnr}, a DFT description of gauge 
enhancing
in circle compactification at self dual radius $R_0$ was provided. There, it 
is shown that enhancing from $U(1)\times U(1)$ to $SU(2)\times SU(2)$ requires 
a dependence of the fields on the internal coordinates $y,\tilde y$ associated 
to a double circle, as we indeed have here. But it also requires an extension 
of 
the tangent space, leading to an $O(d+1+2,d+1+2)$ structure,  that accommodates 
the extra 
massless vector fields associated to winding modes.
The computation was performed  at  $R=\tilde R=R_0$ by keeping only massless 
states,
and it could be extended to $R-\tilde R=R_0\epsilon$ by keeping small masses.
If we tried to generalize  in this direction the procedure described in the 
previous sections, namely
by including   states with $N, \bar N=0,\pm 1$ and keeping GKK massive modes, 
we 
would immediately
run  into trouble.  Since the gauge group is enhanced, now the 
massive 
states (massive gravitons, two-forms, vectors and scalars)  must transform 
under 
 $SU(2)\times SU(2)$. 
However,  there are not enough states, for a given mass,  to fill up these 
representations.
This is again an indication that new fields are needed.
Actually, a string theory analysis, for instance by considering 
the OPE of $SU(2)$ currents with massive gravitons (with $N=\bar N=1$), shows 
that for masses $M^2= 2m \ap$,  gravitons organize into $(2,2), (3,3),\dots 
(m+1,m+1)$ 
representations. In order to fill up these representations, higher 
spin fields are required, which are not contained in the present version of DFT.
Again, the presence of higher order tensors is claimed for, now from gauge 
invariance.

Massive
particles with spin larger than 2 would also be needed if higher powers of 
momentum were considered.
Actually, the three-point functions presented in the Appendix contain higher 
powers of momentum that we have
not included since they go beyond the aim of this paper. However, these higher 
order terms lead to
 higher derivative contributions to the effective action which 
 would of course be necessary if quantum corrections were considered.
In particular, the inclusion of  higher order terms in 
curvature invariants is known to demand the addition of massive tensors  in 
order to fix the short-distance violations of causality 
 \cite{jjm}, and the Regge behavior  required for the resolution of the 
causality problem 
 \cite{vdv}  also calls for higher order tensors in DFT.

Certainly, the effective theory we have constructed does not work as a 
fundamental theory. 
 Nevertheless,
despite the absence of essential ingredients for  full consistency, 
it might  be appealing  by itself.
It encodes  an effective gauge invariant theory with a massless 
sector containing gravity, antisymmetric tensor plus gauge bosons and scalars 
coupled  to towers of GKK massive modes. It is interesting to notice that, 
even if a given  field has a zero mode, it spreads out into  towers of momenta 
and windings.
The simplest case of  a non-zero graviton mass is an interesting
theoretical possibility since it was not until recently that a consistent 
non-linear
theory of massive gravity  could be constructed \cite{drg}.

Even in this simple toroidal scenario it could be interesting 
to look at possible phenomenological consequences and to  explore them in more 
detail.
This aspect is beyond the scope of the present work but let us signal some new 
features that could be worth exploring. 
Many scenarios including KK excitations have been proposed in the literature 
for different physical models.
These proposals deserve being reconsidered in this GKK scenario including 
windings as well as other fields.
On the one hand new fields, associated to antisymmetric tensor 
and dilaton, can be present. Also a new energy scale is built in. 
In fact, even at the circle level two different energy 
scales $\lambda_{KK}=1/R$ and  $\lambda_{windings}= 1/ {\tilde R}$ appear now 
which can 
lead to relevant physical consequences. 

For instance, the type  of models proposed in 
\cite{Han:1998sg}  in the  large  extra dimensions  scenario of \cite{led}
appear to be drastically modified. There,  toroidal bulk KK 
gravity modes were  coupled to Standard Model fields with radii 
$\lambda_{KK} \lesssim M_{string}\sim TeV$. 
However now,  besides the fact that other  fields are present,  the 
$\lambda_{windings}$ 
energy scale will also be present. 
Leaving aside stringy gauge symmetry enhancing, $R={\tilde R}$ self 
dual point situations,  where both windings and KK modes contribute on  the 
same 
footing, are also possible.

KK universal scenarios for dark matter \cite{dmkk} have been extensively 
discussed. 
The consistent incorporation of massive antisymmetric tensors coupled to 
Einstein gravity plus other massless and massive fields could be also appealing 
in this context (see for example \cite{cosmobfield}).
More complex situations, that would require generalizations of this simpler 
toroidal case, provide attractive candidates for dark matter 
\cite{Agashe:2004ci}.
Phenomenology of massive KK gravitons at the  LHC was recently discussed in
\cite{Alvarez:2016ljl},   composite Higgs models associated to bulk KK modes
have been considered in \cite{Alvarez:2010js}, etc.


The ideas developed here could in principle be extended to GKK reductions in 
which
the starting theory has non Abelian gauge fields already in higher 
dimensions (e.g. the heterotic string). These  are just plausible roads 
of research that call for  careful study.

\section*{Acknowledgments}
We thank  E. Andr\'es, P. C\'amara, L. da Rold, D. Marqu\'es,  A. Rosabal and G. 
Torroba for useful 
discussions and comments.
This work was partially supported
by CONICET and  PICT-2012-513.
G. A. thanks the Instituto de F\' isica Te\'orica (IFT
UAM-CSIC) in Madrid for its support via the Centro de Excelencia Severo Ochoa
Program under Grant SEV-2012-0249. G.A. and C.N. thank the A.S.ICTP
for  hospitality and partial support during the completion of this work.
\appendix
\section{Appendix}
\subsection{Extra terms in the DFT action}
\label{ap:Extra terms in the DFT action}

In the original frame formulation of DFT by Siegel 
\cite{Siegel:1993xq,Siegel:1993th}. the action  contains extra terms that are not contained 
in (\ref{action}). Up to total derivatives those can be recast as 
\cite{Geissbuhler:2013uka}
\begin{equation}
\Delta S\, =\, \int d^{2D}X\, e^{-2d}\,\left[\frac12 (S_{\bar A\bar 
B}-\eta_{\bar A\bar B})\eta^{PQ}\, \partial_M E^{\bar A}{}_P\partial^M E^{\bar 
B}{}_Q\, + \, 4\partial_M d\partial^M d \, - \, 4\partial_M\partial^M 
d\right]\, .\label{deltas}
\end{equation}
Here we show that these terms vanish once the level matching 
condition (\ref{lmc}) is imposed.

To show the vanishing of the term proportional to $S_{\bar A\bar B}$ we 
consider 
the following integral
\begin{equation}
I_1=\int d^{2D}X\, \partial^M\partial_M 
\left(e^{-2d}\eta^{PQ}\mathcal{H}_{PQ}\right)=\int d^{2D}X\, 
\partial^M\partial_M \left(e^{-2d}\eta^{PQ}S_{\bar A\bar B}E^{\bar 
A}{}_{P}E^{\bar B}{}_Q\right)=0\, .
\end{equation}
A little of algebra, making use of the property $\mathcal{H}_{PQ}\eta^{PQ}=0$, 
shows that
\begin{equation}
I_1 = 2\int d^{2D}X\, e^{-2d}\eta^{PQ}S_{\bar A\bar B}\left(\partial_M E^{\bar 
A}{}_P\partial^M E^{\bar B}{}_Q + E^{\bar A}{}_P\partial_M \partial^M E^{\bar 
B}{}_Q\right)\, .
\end{equation}
Similarly, for the term in (\ref{deltas}) proportional to $\eta_{\bar A\bar B}$ 
we consider the integral
\begin{equation}
I_2=\int d^{2D}X\, \partial^M\partial_M 
\left(e^{-2d}\eta^{PQ}\eta_{PQ}\right)=\int d^{2D}X\, \partial^M\partial_M 
\left(e^{-2d}\eta^{PQ}\eta_{\bar A\bar B}E^{\bar A}{}_{P}E^{\bar 
B}{}_Q\right)=0\, ,
\end{equation}
that can be recast as
\begin{equation}
I_2=2\int d^{2D}X\, e^{-2d}\eta^{PQ}\left(-\eta_{PQ}\partial_M\partial^M 
d+\eta_{\bar A\bar B}\partial_M E^{\bar A}{}_P\partial^M E^{\bar 
B}{}_Q+\eta_{\bar A\bar B} E^{\bar A}{}_P\partial_M\partial^M E^{\bar 
B}{}_Q\right)\, .
\end{equation}
Finally, for the term proportional to $\partial_M d\partial^M d$, we consider 
the 
integral
\begin{equation}
I_3=\int d^{2D}X\, \partial^M\partial_M e^{-2d} = 2\int d^{2D}X\, 
e^{-2d}\left(2\partial_M d\partial^M d - \partial_M\partial^M d\right)=0\, .
\end{equation}
From $I_1$, $I_2$ and $I_3$, we can therefore 
express $\Delta S$ as
\begin{equation}
\Delta S\, =\, \int d^{2D}X\, e^{-2d}\,\left[-\frac12 (S_{\bar A\bar 
B}-\eta_{\bar A\bar B})\eta^{PQ}\,  E^{\bar A}{}_P\partial_M\partial^M E^{\bar 
B}{}_Q\, - \,(2+D)\partial_M\partial^M d\right]\, .
\end{equation}
And therefore, transforming into momentum space and imposing the level-matching 
condition (\ref{lmc}), we get $\Delta S=0$.
\subsection{String computations}
\label{ap:String computations}
The  results of three-point scattering amplitudes in bosonic string theory 
are  presented here for the case of one compact dimension on a circle of radius $R$. 
They are computed with the vertex operators defined in (\ref{vop}).
We first collect  the amplitudes involving only  massless states and then the 
ones containing at least one  massive state. 
We use a shorthand notation with $h,b,\phi, A, \tilde A$ denoting graviton, 
antisymmetric tensor, scalar and vector fields. Recall that no massive vectors or scalars appear in the circle compactification
and the massive fields are only $h$ and 
$b$.
Dots indicate contractions with Minkowski space-time metric $\eta_{\mu\nu}$.

\subsubsection*{3-point amplitudes for massless states}
\begin{equation}
\begin{aligned}
\langle \Phi\Phi h \rangle &= -(\pi g_{c})\frac{1}{2}\Phi\Phi (k_{1}\cdot\epsilon^{h}\cdot k_{2}) \\
\langle hhh \rangle &=-(\pi g_{c})\frac{1}{2}((k_{2}\cdot\epsilon^{h}_{1}\cdot\epsilon^{h}_{3}\cdot
\epsilon^{h}_{2}\cdot k_{3})+(k_{3}\cdot\epsilon^{h}_{1}\cdot\epsilon^{h}_{2}\cdot\epsilon^{h}_{3}
\cdot k_{2})+(k_{3}\cdot\epsilon^{h}_{2}\cdot\epsilon^{h}_{1}\cdot\epsilon^{h}_{3}\cdot k_{1})\\
& \,\,\,\,\,\, -\frac{1}{2}(k_{3}\cdot\epsilon^{h}_{1}\cdot k_{2})Tr(\epsilon^{h}_{2}
\epsilon^{h}_{3}) -\frac{1}{2}(k_{3}\cdot\epsilon^{h}_{2}\cdot k_{1})Tr(\epsilon^{h}_{1}
\epsilon^{h}_{3})-\frac{1}{2}(k_{1}\cdot\epsilon_{3}\cdot k_{2})Tr(\epsilon^{h}_{1}\epsilon^{h}_{2}) ) \\
\langle AA\Phi \rangle &= (\pi g_{c})\Phi (k_{2}\cdot\epsilon_{1})(k_{1}\cdot\epsilon_{2})\\
\langle \tilde{A}\tilde{A}\Phi \rangle &= -(\pi g_{c})\Phi (k_{2}\cdot\epsilon_{1})(k_{1}\cdot\epsilon_{2})\\
\langle AAh \rangle &=(\pi g_{c})\left((\epsilon_{1}\cdot\epsilon_{2})(k_{1}\cdot\epsilon^{h}\cdot k_{1})
+ (k_{1}\cdot\epsilon^{h}\cdot\epsilon_{2})(\epsilon_{1}\cdot k_{2})+(k_{2}\cdot\epsilon^{h}\cdot\epsilon_{1})
(\epsilon_{2}\cdot k_{1}) \right) \\
\langle A\tilde{A}b \rangle &= (\pi g_{c})\left( (k_{1}\cdot\epsilon^{b}\cdot\epsilon_{2})
(\epsilon_{1}\cdot k_{2}) +(k_{2}\cdot\epsilon^{b}\cdot\epsilon_{1})(\epsilon_{2}\cdot k_{1})
\right) \\
\langle bbh\rangle &=(\pi g_{c})\frac{1}{2}\big( \frac{1}{2}Tr(\epsilon^{b}_{1}\cdot
\epsilon^{b}_{2})(k_{1}\cdot\epsilon^{h}_{3}\cdot k_{2}) + (k_{1}\cdot\epsilon^{b}_{2}\cdot\epsilon^{b}_{1}\cdot
\epsilon^{h}_{3}\cdot k_{1}) + (k_{2}\cdot\epsilon^{b}_{1}\cdot\epsilon^{h}_{3}\cdot\epsilon^{b}_{2}
\cdot k_{3}) \big) \\
\end{aligned}
\end{equation}  
 
\subsubsection*{3-point amplitudes with at least one massive state}
\begin{equation}
\begin{aligned}
\langle hhA \rangle &=(\pi g_{c})\frac{p_{1}}{R}
\left((k_{1}\cdot\epsilon_{3})Tr(\epsilon^{h}_{1}\cdot\epsilon^{h}_{2}) + 
(\epsilon_{3}\cdot\epsilon^{h}_{2}\cdot\epsilon^{h}_{1}\cdot k_{2}) -(\epsilon_{3}
\cdot\epsilon^{h}_{1}\cdot\epsilon^{h}_{2}\cdot k_{1}) \right) \\
\langle hh\tilde{A}\rangle &=(\pi g_{c})\frac{\tilde{p}_{1}}{\tilde{R}}
\left((k_{1}\cdot\epsilon_{3})Tr(\epsilon^{h}_{1}\cdot\epsilon^{h}_{2}) + (\epsilon_{3}
\cdot\epsilon^{h}_{2}\cdot\epsilon^{h}_{1}\cdot k_{2}) -(\epsilon_{3}\cdot\epsilon^{h}_{1}
\cdot\epsilon^{h}_{2}\cdot k_{1}) \right) \\
\langle hh\Phi \rangle &=(\pi g_{c})\frac{1}{2}\Phi Tr(\epsilon^{h}_{1}
\cdot\epsilon^{h}_{2})k_{1L}k_{1R} \\
\langle bb\Phi\rangle &=-(\pi g_{c})\frac{1}{2}\Phi Tr(\epsilon^{b}_{1}\cdot
\epsilon^{b}_{2})k_{1L}k_{1R} \\
\langle bbA \rangle &=-(\pi g_{c})\frac{p_{1}}{R}\left((\epsilon_{3}\cdot k_{1})
Tr(\epsilon^{b}_{1}\cdot\epsilon^{b}_{2}) + (k_{2}\cdot\epsilon^{b}_{1}\cdot\epsilon^{b}_{2}\cdot
\epsilon_{3}) - (k_{1}\cdot\epsilon^{b}_{2}\cdot\epsilon^{b}_{1}\cdot\epsilon_{3}) \right) \\
\langle bb\tilde{A} \rangle &=-(\pi g_{c})\frac{\tilde{p_{1}}}{\tilde{R}}
\left((\epsilon_{3}\cdot k_{1})Tr(\epsilon^{b}_{1}\cdot\epsilon^{b}_{2}) + (k_{2}\cdot
\epsilon^{b}_{1}\cdot\epsilon^{b}_{2}\cdot\epsilon_{3}) - (k_{1}\cdot\epsilon^{b}_{2}\cdot
\epsilon^{b}_{1}\cdot\epsilon_{3}) \right) \\
\langle hbA \rangle &=(\pi g_{c})\left(\frac{p_{1}}{R}(k_{2}\cdot\epsilon^{h}_{1}
\cdot\epsilon^{b}_{2}\cdot\epsilon_{3}) + \frac{p_{2}}{R}(\epsilon_{3}\cdot\epsilon^{h}_{1}
\cdot\epsilon^{b}_{2}\cdot k_{1}) \right) \\
\langle hb\tilde{A} \rangle &=(\pi g_{c})\left(\frac{\tilde{p}_{1}}{\tilde{R}}(k_{2}
\cdot\epsilon^{h}_{1}\cdot\epsilon^{b}_{2}\cdot\epsilon_{3}) + \frac{\tilde{p}_{2}}{\tilde{R}}
(\epsilon_{3}\cdot\epsilon^{h}_{1}\cdot\epsilon^{b}_{2}\cdot k_{1}) \right) 
\end{aligned}
\end{equation}

where $k_{L}=\frac{p}{R} + \frac{\tilde{p}}{\tilde{R}}$ and $k_{R}=\frac{p}{R} - 
\frac{\tilde{p}}{\tilde{R}}$. 

\subsection{Algebra of diffeomorphisms }
\label{ap:global algebra}
Following the discussion  in \cite{dolan,KK}, we can associate  a global infinite parameter algebra to
the infinite modes  $ {{\xi}}^{{\cal P} (\mathbb{M})}(x)$  of
 the GKK expansion of the parameters of local transformations,  in much the same way as  a global Poincar\'e algebra
is associated to general coordinate 
transformations. From
\bea
    {{\xi} }^{\cal P}(x,\mathbb{Y})=\sum_{\mathbb{M}}{}^{\prime}
   {{\xi}}^{{\cal P} (\mathbb{M})}(x)e^{i \mathbb{M}. \mathbb{Y}}\, ,
 \label{paramexpansion}
 \eea
with ${\cal P}=({\rho}, L$), we restrict to 
\bea
 {{\xi}}^{{\rho}(\mathbb{M})}(x)={{a}}^{{\rho}(\mathbb{M})}+
 \omega^{(\mathbb{M}){\rho}}{}_{\nu}x^{\nu}\, ,\\
 {{\xi}}^{{L}(\mathbb{M})}(x)=C^{{L}(\mathbb{M})}\, ,
\eea
where ${{a}}^{{\rho}(\mathbb{M})}, \omega^{(\mathbb{M}){\rho}}{}_{\nu},C^{{L}(\mathbb{M})}$
are constants. The  corresponding generators are
\bea
 \hat 
P_{\rho}^{(\mathbb{M})}&=&ie^{i \mathbb{M}\cdot \mathbb{Y}}\partial_{\rho} \, ,\\
 \hat M_{\mu\nu}^{(\mathbb{M})}&=& e^{i \mathbb{M}\cdot\mathbb{Y}} (
x_{\mu}\partial_{\nu}-x_{\nu}\partial_{\mu})\, ,\\
\hat 
Q_{L}^{(\mathbb{M})}&=&ie^{i \mathbb{M}\cdot\mathbb{Y}}\partial_{L}\, .
 \label{operators}
\eea

 It is easy to check that these operators generate an algebra that corresponds 
to the direct generalization of the algebra found in \cite{dolan}. Namely,
\bea\nn
[ \hat  M_{\mu\nu}^{(\mathbb{M})}, \hat M_{\rho\sigma}^{(\mathbb{N})} 
]&=&i[\eta_{\nu\rho}  \hat  M_{\mu\sigma}^{(\mathbb{M}+\mathbb{N})}+
\eta_{\mu\sigma}  \hat  M_{\nu\rho}^{(\mathbb{M}+\mathbb{N})}
-\eta_{\mu\rho}  \hat  M_{\nu\sigma}^{(\mathbb{M}+\mathbb{N})}
-\eta_{\nu\sigma}  \hat  M_{\mu\rho}^{(\mathbb{M}+\mathbb{N})}]
\\\nn
{[ \hat  M_{\mu\nu}^{(\mathbb{M})},P_{\lambda}^{(\mathbb{N})}]}&=&
i[\eta_{\lambda\nu}P_{\mu}^{(\mathbb{M}+\mathbb{N})}-\eta_{\lambda\mu}P_{\nu}^{(\mathbb{M}+\mathbb{N})}]\\\nn
{[P_{\rho}^{(\mathbb{M})},P_{\mu}^{(\mathbb{N})}]}&=&0\\\nn
{[Q_{L}^{(\mathbb{M})},  \hat M_{\mu\nu}^{(\mathbb{N})}]}&=&-\mathbb{
N}_L 
M_{\mu\nu}^{(\mathbb{M}+\mathbb{N})}\\\nn
{[Q_{L}^{(\mathbb{M})}, P_{\mu}^{(\mathbb{N})}]}&=&-\mathbb{N}_L 
P_{\mu}^{(\mathbb{M}+\mathbb{N})}\\
{[Q_{L}^{(\mathbb{M})},Q_{S}^{(\mathbb{N})}]}&=&-\mathbb{N}_LQ_{S}^{(\mathbb{M}
+\mathbb{N})}+\mathbb{M}_S Q_{L}^{(\mathbb{M}+\mathbb{N})}
\eea

We see that the zero modes lead to the $d$ dimensional Poincar\'e algebra.
Also, from the last equation we notice that, for $L=S$
\bea
{[Q_{L}^{(\mathbb{M})},Q_{L}^{(\mathbb{N})}]}&=&(\mathbb{M}_L-\mathbb{N}_L)
Q_{L}^{(\mathbb{M}
+\mathbb{N})}\, ,
\eea
which is a Virasoro algebra (with no central charge) for each value of 
$L=1,\dots 2n$.

For the case of the circle we would have $\mathbb{M}= (m_1,m_2)=(m,\tilde m)$ 
with $m=0$ or $\tilde m=0$ due to LMC.

Notice that if we choose $\mathbb{M}=(m,0)$  and  $\mathbb{N}=(n,0)$ with 
$m,n=\pm 1, 0$.  Then $ \hat Q_{1}^{(\mathbb{M})}\equiv\hat 
Q_{1}^{(\pm 1)}, Q_{1}^{(0)}$ and $ \hat P_{\mu}^{(0)},  \hat 
M_{\mu\nu}^{(0)}, \hat Q_{2}^{(0)}$ close a {Poincar\'e}$ \otimes SO(1,2)$ 
algebra. In the same way, exchanging $1\leftrightarrow 2$, namely, windings with 
momenta, another $SO(1,2)$ algebra is obtained. Thus,  finally the original 
Poincar\'e algebra is enlarged to { \rm Poincar\'e} $\otimes SO(1,2)^2$.
It was shown in \cite{Salam:1981xd} that, in the circle case in field theory, 
the massive KK states organize into an infinite  dimensional (non-unitary) $R$
representation of $SO(1,2)$. In DFT on the circle, windings and momenta are 
decoupled,  so massive  KK momenta states will fill up the infinite 
dimensional  representation of the first algebra whereas windings will organize in 
a similar representation of the second one, namely $(R,1)+(1,R)$.

In the generic case we can proceed in the same way by choosing the GKK momenta 
at the position $L$,  $\mathbb{M}_L=0,\pm 1$ with all other components 
vanishing. In this case we would have { \rm Poincar\'e} $\otimes SO(1,2)^{2n}$.
Since massive states with $M^2= \mathbb{P}\cdot {\cal M} \cdot\mathbb{P}$ mix windings and momenta 
the analysis of representations is more involved and we will not perform 
it in the present  work. 

Even if the above algebra is a symmetry of the original Lagrangian, it is broken to
Poincar\'e $\times U(1)^n\times U(1)^n$ by the vacuum (\ref{vacuum}). This can be 
easily verified  by inserting the mode expansions (\ref{paramexpansion}) 
to compute the transformations of the fields
 $g_{\mu \nu}, A^{M}_{\mu}, b_{\mu \nu},{\cal H}_{MN }$ and by requiring the vacuum (\ref{vacuum})
 to be invariant under these transformations. 
$ {{\xi}}^{{\cal P} (\mathbb{M})}$, with $ \mathbb{M}\ne 0$ correspond to
broken generators associated to Goldstone bosons.

\end{document}